# Surface composition and properties of Ganymede: updates from ground-based observations with the near-infrared imaging spectrometer SINFONI/VLT/ESO.

**Ligier N. [1], Paranicas C. [2], Carter J. [3], Poulet F. [3], Calvin W.M. [4], Nordheim T.A. [5], Snodgrass C. [1,6], Ferellec L. [3]**

[1]School of Physical Sciences, The Open University, Milton Keynes MK7 6AA, UK. [2]Applied Physics Laboratory, Johns Hopkins University, Laurel, MD 20723, USA. [3]Institut d'Astrophysique Spatiale, Université Paris-Saclay, 91405 Orsay Cedex, France. [4]Geological Sciences & Engineering, University of Nevada, Reno, NV 89557, USA. [5]Jet Propulsion Laboratory, Pasadena, CA 91109, USA. [6]School of Physics and Astronomy, University of Edinburgh, Edinburgh EH8 9YL, UK.

First author's e-mail: nicolas.ligier@open.ac.uk

# ABSTRACT

Ganymede's surface exhibits great geological diversity, with old dark terrains, expressed through the surface composition, which is known to be dominated by two constituents: $H_2O$-ice and an unidentified darkening agent. In this paper, new investigations of the composition of Ganymede's surface at global scale are presented. The analyses are derived from the linear spectral modeling of a high spectral resolution dataset, acquired with the near-infrared (1.40 – 2.50 µm) ground-based integral field spectrometer SINFONI (**SIN**gle **F**aint **O**bject **N**ear-IR **I**nvestigation) of the Very Large Telescope (VLT hereafter) located in Chile. We show that, unlike the neighboring moon Europa, photometric corrections cannot be performed using a simple Lambertian model. However, we find that the Oren-Nayar (1994) model, generalizing the Lambert's law for rough surfaces, produces excellent results. Spectral modeling confirms that Ganymede's surface composition is dominated by $H_2O$-ice, which is predominantly crystalline, as well as a darkening agent, but it also clearly highlights the necessity of secondary species to better fit the measurements: sulfuric acid hydrate and salts, likely sulfates and chlorinated. A latitudinal gradient and a hemispherical dichotomy are the strongest spatial patterns observed for the darkening agent, the $H_2O$-ice, and the sulfuric acid: the darkening agent is by far the major compound at the equator and mid-latitudes ($\leq \pm 35°N$), especially on the trailing hemisphere, while the $H_2O$-ice and the sulfuric acid are mostly located at high latitudes and on the leading hemisphere. This anti-correlation is likely a consequence of the bombardment of the constituents in the Jovian magnetosphere which are much more intense at latitudes higher than $\pm 35°N$. Furthermore, the modeling confirms that polar caps are enriched in small, fresh, $H_2O$-ice grains (i.e. $\leq 50$ µm) while equatorial regions are mostly composed of larger grains (i.e. $\geq 200$ µm, up to 1 mm). Finally, the spatial distribution of the salts is neither related to the Jovian magnetospheric bombardment nor the craters. These species are mostly detected on bright grooved terrains surrounding darker areas. Endogenous processes, such as freezing of upwelling fluids going through the ice shell, may explain this distribution. In addition, a small spectral residue that might be related to brines and/or hydrated silica-bearing minerals are located in the same areas.

# INTRODUCTION

Ganymede, the largest satellite of the Solar System with its radius of 2631 km, is one of the four Galilean satellites of the Jovian system. Ganymede holds an important place in the Solar System in terms of scientific interest, especially in geology. Indeed, some areas on its surface identified as bright grooved terrains (or *sulci*, singular *sulcus*, Latin for groove) attest to the existence of past endogenic activity such as plate tectonics and cryogenic volcanism (Pappalardo et al. 1998, Schenk et al. 2001, Pappalardo et al. 2004). Ganymede was also discovered by the Galileo mission to be the first and thus far only satellite to have its own intrinsic magnetic field (Kivelson et al. 1996). The analysis of its global density with the spherical harmonics of its gravitational field undoubtedly shows that Ganymede is a differentiated body composed of three layers (Anderson et al. 1996): a liquid or partially liquid metallic core (Schubert et al. 1996), surrounded by a silicate mantle which is covered by an ~800 km thick ice shell. Based on a tidal heating model derived from Galileo's magnetometer data, a ~200 km thick water layer partially or entirely liquid is expected, sandwiched between two layers of ice, at a depth of ~150 km (Kivelson et al. 2002, Sotin & Tobie 2004).

Ganymede's liquid layer is assumed to be enriched in ions, especially $Na^+$, $Mg^{2+}$ and $SO_4^{2-}$, because of the aqueous alteration of carbonaceous chondrites that are considered to represent the bulk material that accreted to form the Galilean moons (Kargel et al. 2000, Fanale et al. 2001). As these assumptions cannot be verified by direct, in-situ investigations, the analysis of the surface composition using remote sensing spectroscopy is the preferred method to determine the physical and chemical properties of this subsurface ocean, with implications for its habitability. To date, compositional information has mainly been obtained by the Near-Infrared Mapping Spectrometer (NIMS) instrument onboard the Galileo spacecraft, covering the 0.7 – 5.2 µm range with a typical spectral resolution of 0.025 µm (Carlson et al. 1992) and a spatial resolution varying from ~80 km/pix to ~2 km/pix (Hibbitts et al. 2003).

Like the majority of giant planet satellites, Ganymede's surface has long been known to be dominated by $H_2O$-ice (Kieffer & Smythe 1974, Pollack et al. 1978, Calvin et al. 1995). According to a study based on the Fresnel reflection peak of the $H_2O$-ice near 3.1 µm, the first few millimeters of Ganymede's $H_2O$-ice are presumed to be both crystalline and amorphous, suggesting approximately equal rates between two processes: (i) thermal crystallization due to thermal cycles and (ii) radiation-induced amorphization resulting from bombardment by charged particles orbiting inside the Jovian magnetosphere (Hansen & McCord 2004). However, because the charged particles are believed to precipitate more heavily onto polar regions, Ganymede's intrinsic magnetic field plays an important role in the crystallization versus amorphization balance. Indeed, Hansen & McCord (2004) globally observed the crystalline form of $H_2O$-ice at low latitudes and its amorphous form at high latitudes (≥ ±40°). In the visible, these high latitudes actually exhibit a much brighter surface than the other areas (except fresh craters) and are thus commonly named polar caps. A correlation seems clear between the polar cap boundaries and the transition between open and closed magnetic field lines of the moon

(Kivelson et al. 1998, Khurana et al. 2007), probably suggesting an external (charged) weathering agent is responsible. For example, the existence and the brightness of these caps could be attributed to the Jovian magnetospheric plasma or charged particle bombardment and its associated sputter-induced redistribution of $H_2O$ molecules (Johnson 1997, Khurana et al. 2007, Plainaki et al. 2015).

Beside $H_2O$-ice, multiple weak non-$H_2O$ absorptions have been detected in NIMS' spectra. The most prominent of these absorptions is found at ~4.25 μm and has been identified as $CO_2$ frost (McCord et al. 1997, McCord et al. 1998a, Hibbits et al. 2003). The $CO_2$, which is likely to be hosted in non-$H_2O$ materials (McCord et al. 1998a), is undetected in polar regions, while dark terrains and other areas are locally $CO_2$-rich; this distribution appears to be mainly controlled by geological processes (Hibbits et al. 2003, Pappalardo et al. 2004). Other weak absorptions are located at (i) 3.4 μm, (ii) 3.88 μm, (iii) 4.05 μm and (iv) 4.57 μm (McCord et al. 1997, McCord et al. 1998a). These absorptions are respectively suggested related to (i) CH-bearing compounds such as hydrocarbons, (ii) SH-bearing compounds such as $H_2S$, (iii) $SO_2$ and "altered" sulfur materials, and (iv) tholins or sulfate salts and other sulfur-oxygen minerals (McCord et al. 1998a, Domingue et al. 1998). Some of these species may originate from outside the Jovian system and may exist as small inclusions in the surface material. Because the Jovian magnetosphere delivers intense radiation bombardment of the surface of the Galilean moons, the C-bearing species are likely destroyed by radiolysis and could possibly be decomposed into pure carbon, thus playing the role of a spectrally flat darkening agent (McCord et al. 1997, McCord et al. 1998a).

The existence of such darkening, or "grey", non-icy contaminant on the Galilean moons, especially Ganymede and Callisto, has been discussed for almost 50 years beginning with telescopic observations (Pilcher et al. 1972, Lebofsky 1977, Pollack et al. 1978) but it remains unidentified (Stephan et al. 2017). In addition to the radiolysis of carbonaceous species, hydroxylated or hydrated silicates and salts are also good candidates for the unknown darkening agent (Pilcher et al. 1972, Calvin & Clark 1991, Calvin et al. 1995, McCord et al. 2001b, Pappalardo et al. 2004). While it is widely accepted that hydrated silicate materials should likely be the result of aqueous alteration of endogenous or impact-emplaced materials, the origin of the salts, likely S-bearing, is debated. They could be the product of Ganymede's surface bombardment by exogenous sulfur ions (Strazzulla et al. 2009, Ding et al. 2013), but some could also originate from Ganymede's subsurface ocean (McCord et al. 2001b). Carbonaceous asteroid deposits have been proposed as analogs for the contaminant (Calvin & Clark 1991, Schenk & McKinnon 1991). Whatever the origin of this darkening agent, it is generally accepted that it should be a fundamental component widespread over the entire surface of Ganymede (Helfenstein et al. 1997, Pappalardo et al. 2004).

Most of these results are from NASA's Galileo mission, which ended in 2003. Important technological advances have been made since then. Moreover, many questions about Ganymede's surface composition remain unanswered. For these reasons, Ganymede is a major target of interest, and is a primary target of ESA's first Cosmic Vision L-class mission, the Jupiter Icy moons Explorer, or JUICE, with a planned launch in 2022, and arrival in 2029. JUICE is scheduled to orbit around

Ganymede during the last phase of the mission. The chemical and physical properties of Ganymede's surface will be investigated with different instruments. One of them is the Moons And Jupiter Imaging Spectrometer (MAJIS), an imaging spectrometer working in the visible and near-infrared wavelength domain (Langevin et al. 2014). In order to prepare the radiometric budget of this instrument and to bring added value to previous NIMS results, a ground-based observation campaign was led from October 2012 to March 2015 with the integral field spectrometer SINFONI, an instrument mounted on the Yepun telescope (UT4) of the VLT in Chile (Eisenhauer et al. 2003). The exceptionally high spatial and spectral sampling of this instrument (see the next section), coupled with its performing adaptive optics and its excellent signal-to-noise ratio (SNR), allow the potential detection and mapping of narrow spectral absorptions as has already been shown in a recently published paper about the surface composition of Europa (Ligier et al. 2016).

This paper is organized as follows: (i) we first present the instrument, the observations and the reduction process developed in this study; (ii) the second part is dedicated to an overall analysis of some selected spectra and their characteristics; (iii) then spectral modeling is performed with a linear unmixing model using a dark end-member and cryogenic reference spectra of $H_2O$-ice and hydrated species; (iv) finally, the presence, spatial distribution and origin of these species are discussed.

## INSTRUMENT, OBSERVATIONS AND DATA REDUCTION

### 1. The SINFONI instrument

SINFONI is an integral field spectrometer operating in the near-infrared range, more precisely from 1.05 µm to 2.50 µm, with four different gratings covering the atmospheric windows: J, H, K and the combined H+K band (Eisenhauer et al. 2003). To cover the widest spectral range possible, only the latter was used in this study. The smallest field of view (FoV) was used to maximize the spatial resolution providing 0.8 by 0.8 arcsecond, divided into 64 times 32 pixels of 12.5 by 25 milliarcsecond (mas) size. The FoV was then automatically rebinned into a 64 by 64 pixel image-cube just after the acquisition, leading to an actual spatial sampling of 12.5 by 12.5 mas. A spectrum, made of 2200 points at different wavelengths covering the 1.40 – 2.50 µm range (thus meaning $5 \times 10^{-4}$ µm as spectral sampling), is associated to every pixel of the FoV. Hence, each acquisition results in a cube (x, y, $\lambda$) of data, with dimensions $64 \times 64 \times 2200$.

### 2. Observations

The observation campaign took place from October 2012 to March 2015 during four different nights close to opposition to maximize the spatial resolution. This led to angular diameters of Ganymede at least twice as large as the FoV, i.e. from 1.62 to 1.78 arcsec. To fully cover the satellite's surface, a mosaic of ten overlapping observations was defined (figure 1). Each mosaic was preceded and/or followed by sky background frames and telluric standard star observations. Table 1 provides a log of the four observation blocks, including the main observational parameters (date, exposure time, airmass, Strehl ratio and phase angle). That table also includes geographical and geometrical information that are not only important for the data reduction, but also for the map projection.

### 3. Data reduction

#### 1. *First steps of the data reduction and the mosaic reconstruction*

The data reduction is similar to that described in Ligier et al. (2016). First, the ESO pipeline and the telluric correction presented in sub-sections 2.3.1 and 2.3.2 of Ligier et al. (2016), are the same. While the ESO pipeline mostly consists in (i) removing detector's artifacts, (ii) performing the bias and dark subtraction, (iii) flat-fielding (iv) calibrating the wavelength and (v) correcting the geometric distortion, the telluric correction removes the spectral contribution of the Sun. Then follows the mosaic reconstruction, which is different than the one performed for Europa; insofar as Ganymede's mosaics are composed by ten frames while Europa's are only made of five, there is a more complex distribution of overlapping pixels between each frame (figure 1). To mosaic each SINFONI FoV, the total flux ($F$) from 1.46 µm to 2.41 µm (excluding the range from 1.79 µm to 1.96 µm because of atmospheric contaminations) is summed for all the overlapping pixels. A factor is calculated to normalize every surrounding frame of the mosaic with the first central one (numerated as #1 in table 2):

$$f_1 = \frac{F_{X\ frame}}{F_{1^{st}\ frame}}$$

With $f_1$ ranging from 0.974 to 1.060 (table 2), the flux between different frames of a same observation is quite similar. Moreover, by covering the same area, the last, tenth, frame allows a double-check of the consistency. Finally, we obtain a single 3-dimensional cube for each observation after applying these normalizing factors.

#### 2. *Photometric corrections*

Photometric corrections are needed in order to obtain geometrically corrected reflectance spectra. Although a Lambertian model worked well for Europa, it was quickly apparent that this would not be a good approximation for the surface of Ganymede (figure 2c). Consequently, another photometric correction is applied, namely the qualitative model of Oren-Nayar, generalizing the Lambertian law for rough surfaces (Oren & Nayar 1994). This model, little used compared to some others in planetary sciences, is expressed as follows:

$$L_r(\theta_r, \theta_i, \phi_r - \phi_i, \sigma) = \frac{\rho}{\pi} E_0 \cos \theta_i \times \left(A + B \times Max[0, \cos \phi_r - \phi_i] \times \sin \alpha \times \tan \beta \right)$$

$$\text{with}\ \ A = 1.0 - 0.5 \times \frac{\sigma^2}{\sigma^2 + 0.33}, \ \ B = 0.45 \times \frac{\sigma^2}{\sigma^2 + 0.09}, \ \ \alpha = Max(\theta_i, \theta_r), \ \ \beta = Min(\theta_i, \theta_r)$$

Thus, the radiance $L_r$ only depends on the angles of incidence ($\theta_i, \phi_i$) and reflection ($\theta_r, \phi_r$), the surface roughness ($\sigma$), the albedo of the surface ($\rho$), and the irradiance of a surface illuminated head-on ($E_0$). Note that this model reduces to the Lambert law when $\sigma = 0°$. Furthermore, in our case of Earth-based data acquired near the opposition, phase angles (table 1, Φ column) are low, meaning that angles of incidence and reflection are close, so: $\theta_r \approx \theta_i$ and $\phi_r \approx \phi_i$. Therefore, the model can be simplified:

$$L_r(\theta_i, \sigma) \approx \frac{\rho}{\pi} E_0 \cos\theta_i \times (A + B \sin\theta_i \tan\theta_i)$$

The $\theta_i$ of every 3D-cube's pixel depends on its distance with the sub-solar point ($SSP$) and the radius of the satellite ($r$), here 2631 km. As already shown in Ligier et al. (2016), it is calculated as follows:

$$\theta_i = \cos^{-1}\left(\sqrt{1 - \left(\frac{SSP}{r}\right)^2}\right)$$

Thus, multiple empirical tests varying the $\sigma$ parameter, expressed in $A$ and $B$, were performed to obtain the geometrically normalized reflectance. The consistency of the results was checked visually by verifying that pixels with high inclination angles ($\theta_i \geq 30°$) do not exhibit over-estimated or under-estimated reflectance compared to the rest of the disk (figure 3). An excellent photometric correction is eventually obtained up to $\theta_i = 65°$. Few inclination residuals are observed while large geological units and fresh craters are recovered. The roughness parameters for all the observations are: (i) $\sigma = 21° \pm 6°$ for the 2012/10/30 observation, (ii) $\sigma = 16° \pm 6°$ for the 2012/11/23 observation, (iii) $\sigma = 16° \pm 6°$ for the 2015/02/17 observation, and (iv) $\sigma = 19° \pm 6°$ for the 2015/03/08 observation (figure 2d). Based on the error bars on $\sigma$ values, the reflectance level uncertainty is estimated to ± 20% of the reflectance obtained in the worst cases, i.e. pixels with very high inclination angles ($\theta_i \geq 55°$). Lastly, to complete photometric corrections, a phase angle correction is applied to normalize the flux measured regardless of the phase angle. Using the observations with the lowest phase angle as reference (2012/11/23), and based on phase curves obtained with terrestrial observations in the visible (Domingue & Verbiscer 1997), it leads to correcting factors $f_2 = 1.05$, $f_2 = 1.00$ and $f_2 = 1.04$ respectively for the 2012/10/30, 2015/02/17 and 2015/03/08 observations. Figure 2e shows the final result of the whole data reduction applied until this step.

### 3. *From multiple 3D-cubes to a single 3D-map*

To obtain a single 3D-map for the entire satellite from multiple 3D-cubes, we account for the North pole position angle at the time of each observation (table 1, P.A. column). From that point, each 3D-cube is projected onto a 360 × 180 pixel array using an equirectangular projection. Then, after the projection, the next step consists in merging the observations together, regardless of the atmospheric conditions in which they were acquired. Because of varying observational conditions, some spectral inconsistencies, such as small changes in slope and miscalibration of the reflectance level, may appear between observations. To avoid most of these artifacts, a last normalization process using the best observation selected on the Strehl ratio value given in table 1 (here 2012/11/23) is performed prior to merging. This normalization is composed of three steps: (i) an average spectrum for overlapped pixels belonging to the reference observation is calculated ($S_{ref.}(\lambda)$), (ii) another average spectrum for the same overlapped pixels belonging to a non-reference observation is calculated ($S_{non-ref.}(\lambda)$) and (iii) the whole non-reference observation is finally corrected by the spectral ratio $r(\lambda) = \frac{S_{ref.}(\lambda)}{S_{non-ref.}(\lambda)}$. This procedure will propagate systematic features existing specifically in spectra of the reference observation, but at the same time, all systematic features that only exist in the three non-reference observations, which may be more numerous, will be suppressed; therefore, special attention was paid during spectral analysis to distinguish real spectral features from instrumental, non-physical, ones. From there, the final 3D-map is obtained by merging the four observations. Pixels sampled multiple times are simply averaged. As a result, approximately 80% of Ganymede's surface between -60°N and +60°N of latitude is covered (figure 4a), providing an unprecedented amount of spectral data about Ganymede's surface, all with excellent spectral and spatial sampling (respectively 0.5 nm and $\sim 45 \times 45$ km$^2$).

**MAPS AND SPECTRA**

1. **H$_2$O-ice band depth maps**

For icy bodies, an easy way to look at the surface composition is to produce band depth maps, and particularly those related to H$_2$O-ice features. In H+K spectral range, the most prominent features are located at 1.50 µm, 1.57 µm, 1.65 µm and 2.02 µm. The band depth ($BD$) is defined as follows:

$$BD(\lambda) = 1 - \frac{Band(\lambda)}{Continuum(\lambda)}$$

Consequently, it means the higher the $BD(\lambda)$, the stronger the absorption and possibly the higher the abundance (for a give grain size). For this study, two absorptions were studied: at 2.02 µm and at 1.65 µm. The latter absorption is characteristic of crystalline ice (Grundy & Schmitt 1998). Consequently, while the $BD(2.02)$ map (figure 4b) characterizes the distribution of the H$_2$O-ice in general, the $BD(1.65)$ map (figure 4c) specifically characterizes the distribution of the crystalline form of H$_2$O-ice. To get $BD(\lambda)$, the continuum has to be calculated from measured values: for the 2.02 µm absorption, the left and right shoulders are located at 1.74 µm and 2.24 µm respectively, while they are located at 1.615 µm and 1.680 µm for the 1.65 µm absorption. Lastly, to reduce noise influence, the continuum and band center used to calculate $BD(\lambda)$ correspond to the median reflectance over 11 spectels (i.e. a ± 2.5 nm range). $BD(2.02)$ and $BD(1.65)$ maps highlight a clear correlation (figure 5). At first order, a strong latitudinal gradient is observed on both hemispheres, meaning that the higher the latitude, the stronger the H$_2$O-ice bands. An ice-depleted area is observed on Ganymede's trailing hemisphere, very similar to the bullseye pattern detected on Europa (McEwen 1986, Carlson et al. 2005, Grundy et al. 2007, Ligier et al. 2016). But, unlike Europa, the pattern is not perfectly centered on the apex of the trailing hemisphere [0°N, 270°W] but shifted ~15° eastward. This pattern could suggest a stronger influence of the Jovian magnetosphere at the equator of Ganymede's trailing hemisphere, however it is interesting to note that such hypothesis would be in total disagreement with the total ion flux reaching Ganymede's surface modeled by Fatemi et al. (2016). This precise point will be addressed in details during the discussion. Nevertheless, the bullseye pattern seems to coincide at least partially with Melotte Regio [-12°N, 245°W]. Other geomorphological units and large craters are identifiable on the reflectance and band depth maps: the dark southern part of Galileo Regio [45°N, 127°W], the less-cratered part of Perrine Regio [34°N, 28°W], the western part of Barnard Regio [-7°N, 12°W] and Nicholson Regio [-33°N, 6°W], Uruk sulcus [3°N, 160°W] between Marius Regio [3°N, 188°W] and Galileo Regio, Xibalba sulcus [43°N, 71°W] separating Galileo Regio and Perrine Regio, Hershef crater [47°N, 269°W], Tashmetum crater [-40°N, 265°W], Osiris crater [-38°N, 166°W] with its ejecta and its surrounding craters, Ashîma crater [-39°N, 123°W], Cisti crater [-32°N, 64°W], and finally Tros crater [11°N, 27°W]. All these structures are named and delimited in the USGS global geologic map of Ganymede produced by Collins et al. (2014), and longitudes and latitudes are provided by the website of the International Astronomical Union: https://planetarynames.wr.usgs.gov/.

Despite the strong correlation between the $BD(2.02)$ and $BD(1.65)$ maps, some differences are still visible. Indeed, while the area around the apex of the trailing orbital hemisphere has the lowest value for the $BD(2.02)$ map, the lowest value for the $BD(1.65)$ map is shared with Galileo Regio. More generally, all regiones (plural of regio) mentioned above are proportionally closer to the lowest values of the $BD(1.65)$ map than those of the $BD(2.02)$ map. At the same time, the southern part of the leading orbital hemisphere (60°N, 60°W) has the highest value of the $BD(2.02)$ map, while the highest values of the $BD(1.65)$ map are craters, such as Cisti and especially Osiris with its numerous surrounding craters. Using simple band depths, all regiones exhibit smaller crystalline $H_2O$-ice signatures in contrast to the brighter craters. These trends will be investigated through spectral modeling later in the paper.

## 2. Spectral characteristics and implications

### 1. *General overview*

Figure 6 shows different spectra of Ganymede's surface. As expected, most areas are largely dominated by four $H_2O$-ice absorptions: 1.50 µm, 1.57 µm, 1.65 µm and 2.02 µm. While band centers are more or less located at 1.50 µm, 1.57 µm and 1.65 µm whatever the spectra, the center of the band at 2.02 µm is clearly shifted to shorter wavelengths, up to 1.95 µm in some cases, for $H_2O$-ice poor or relatively poor spectra. It is interesting to note that the right bound of the 2.02 µm absorption, i.e. the peak at 2.24 µm, is also slightly shifted to shorter wavelengths for spectra showing a strong distortion of the 2.02 µm band. That kind of distortion of this ice-related absorption band, also largely observed for Europa's spectra, is commonly attributed to sulfate salts and/or some hydrated form of sulfuric acid (McCord et al. 1998a, Carlson et al. 1999, McCord et al. 2001b, Carlson et al. 2005). There is no other absorption feature in this wavelength range, except the "broken" shoulder at 1.74 µm. Such an abrupt slope break is unusual for $H_2O$-ice and therefore could be related to one or multiple non-$H_2O$ species. Hydrated silicates and carbonaceous compounds, spectrally relevant candidates for the darkening non-icy contaminant on Ganymede's surface (Pilcher et al. 1972, Pappalardo et al. 2004), do not possess any absorption features at this wavelength. However, some salty species have important bands there at cryogenic temperatures, especially hydrated Cl-salts (Hanley et al. 2014). The questions raised by this absorption will be further addressed in this paper, especially through spectral modeling.

### 2. *Ice temperature*

It is widely observed that the infrared bands of $H_2O$-ice are temperature-dependent, especially the 1.65 μm band which is particularly sensitive (Grundy & Schmitt 1998, Leto et al. 2005, Mastrapa et al. 2008). The colder the ice, the shorter the band center's wavelength: more precisely, the band center varies from 1.646 μm to 1.656 μm for temperatures respectively equal to 150K and 80K. With a spectral sampling of $5 \times 10^{-4}$ μm, SINFONI data are sufficiently resolved to investigate the precise location of the band center. To do so, each spectrum is recalculated by integrating spectra of surrounding pixels inside a 5-pixel diameter circle. This size is chosen because it corresponds to SINFONI's Airy spot at 2.0 μm, which is approximately 60 mas. Then, following the same methodology described in the sub-section 3.2.2 of Ligier et al. (2016), we obtain a band center located at 1.6495 ± 0.0035 μm for the fresh Osiris crater, the most ice-rich spectra of the figure 6; Mastrapa et al. (2008) give an ice temperature of ~125 ± 25K for such a band center. To test the representativeness of this temperature over the whole surface, the method is applied to every pixel: approximately 99% of pixels have a band center in the 1.6475 – 1.6515 μm range, while 51% of those pixels have band centers located at 1.6495 μm. Taking into account the standard deviation on the position, i.e. ± 0.0035 μm for most cases, a temperature in the 90 – 160K range at Ganymede's surface is obtained. This temperature range is fully consistent with Galileo Photopolarimeter-Radiometer (PPR) measurements (Orton et al. 1996), but the mapping of the position of this band center is different from what was observed by the PPR sensor (figure 7): as expected, one observes a clear leading/trailing dichotomy suggesting the influence of the Jovian magnetosphere on the 1.65 μm position. It suggests that ice contamination by non-$H_2O$ material and/or intimate mixtures lead to the spectral shift of this absorption. This result is in agreement with what has been recently observed in Europa's spectra (Ligier et al. 2016). Hence, the spectra of the most ice-rich pixels, i.e. those corresponding to fresh craters or to the leading hemisphere, are considered to be better proxy to determine Ganymede's surface temperature, ~125 ± 25K.

## GANYMEDE'S SURFACE LINEAR MODELING

### 1. Methodology

#### 1. *The linear spectral modeling*

After studying the spectral diversity of Ganymede's surface qualitatively, the next step consists in getting quantitative estimates of the surface composition through linear spectral modeling. Linear modeling is a linear deconvolution technique based on the principle that each measured spectrum is the result of the combination of laboratory reference spectra acquired at a temperature representative of the studied surface, in this case ~125 ± 25K. The following formula is used to calculate the model:

$$M(\lambda) = \sum_{i=1}^{N} A_i \times L_i(\lambda) \times S$$

A nonlinear, e.g. radiative transfer, approach for the modeling is generally preferred insofar as such approach produces better quantitative results and could help to improve the fits further. However, the radiative transfer approach requires optical constants for each end-member, which are currently critically lacking at cryogenic temperatures except for crystalline and amorphous $H_2O$-ice. As a consequence, the modeled spectrum $M(\lambda)$ corresponds to the sum of $N$ laboratory spectra $L_i(\lambda)$ on which a slope parameter $S$ and varying abundance coefficients $A_i$ are applied. For this study, $M(\lambda)$ is obtained using an automated iterative method based on the downhill simplex method described in Nelder et al. (1965). The method provides iterative calculations by varying $S$ and $A_i$ to obtain the lowest total root mean square value ($RMS_{TOT}$). $RMS_{TOT}$ is the sum of four $RMS_i$ calculated in four different wavelength domains: $RMS_1$ for the 1.46 – 1.79 μm range, $RMS_2$ for the 1.94 – 2.41 μm range, $RMS_3$ for the 1.56 – 1.67 μm range and $RMS_4$ for the 1.97 – 2.15 μm range. $RMS_3$ and $RMS_4$ play an important role to fit as much as possible the main absorptions in the spectra (figure 4). The modeled spectrum is saved once modeling stops improving $RMS_{TOT}$ at a $10^{-3}$ level after few tens of additional simulations.

## 2. *The laboratory spectra*

To be representative of the surface conditions, only cryogenic spectra acquired from 80K up to 160K were considered as relevant for the spectral reference library. As the published laboratory spectra are mostly acquired at ambient or Martian temperature, cryogenic laboratory spectra are relatively uncommon. Therefore, only very few spectra are appropriate for spectral modeling of icy satellite surfaces. These laboratory spectra can be separated into three types: (i) $H_2O$-ice, amorphous and crystalline, (ii) hydrated sulfuric acid, and (iii) salts, such as sulfates, carbonates and chlorinated. These minerals are expected to be present on the surface of Ganymede based on the multiple published spectral modeling studies mentioned in the introduction. The reflectance spectra of amorphous and crystalline ice are the only reference spectra in our cryogenic library that were derived from the Shkuratov radiative transfer model (Shkuratov et al. 1999, Poulet et al. 2002) using the optical constants of both types of ice provided by Mastrapa et al. (2008) and the GhoSST (**G**renoble astro**ph**ysics and planet**o**logy **S**olid **S**pectroscopy and **T**hermodynamics) database (https://ghosst.osug.fr), respectively. Finally, another component is required for modeling: a darkening agent, spectrally flat over the entire wavelength range. This artificial end-member, mentioned in the introduction, actually plays the role of the darkening, non-icy, material in order to bring the modeled albedo levels to match those that are observed. Constraints on this dark material have to be applied though. This question will be investigated and discussed through spectral modeling. All the spectra and their associated acquisition parameters are summarized table 3.

## 3. *Laboratory spectra selection for modeling*

As the main objective of this study is to get a global overview of the chemical properties across Ganymede's surface, the spectral modeling has to be performed on every pixel to obtain abundance maps. These maps will provide information about the spatial distribution of the different species used in the model. To determine the relevant reference spectra to use as inputs for the global spectral modeling, a dozen representative spectra, as shown in the figure 6, are modeled with different combinations of the laboratory spectra used as inputs. An iterative approach was chosen for the spectral unmixing, starting from a small set of laboratory spectra. Steps of this selection process are presented in figure 8 (split in two parts to ease reading) and are explained in the following paragraph. Table 4 provides the abundances of best-fit modeled spectra at each step.

The spectra were first modeled only with $H_2O$-ice (crystalline and amorphous with all their grain sizes) and the darkening agent with its reflectance level varying between 0.15 and 0.30. This model is referred as "Model #1" (M1 hereafter) in figure 8. Even if some models could be considered consistent at NIMS/Galileo spectral resolution, SINFONI's spectral resolution highlights discrepancies with the measurements. Surprisingly, these discrepancies are stronger for spectra with strong $H_2O$-ice related absorptions, except for the most icy spectrum [-42°N, 172°W] which is already quite consistent. Overall, the left part of the 2.02 µm absorption and the shoulder at 1.74 µm are poorly modeled, and a majority of models underestimate the reflectance level in the 1.51 – 1.56 µm and 2.30 – 2.41 µm ranges. To improve the fits, either the sulfuric acid hydrate $H_2SO_4.8(H_2O)$ or the twelve salts (five sulfates, one carbonate, six chlorinated) available in table 3 were added in inputs. These models are respectively referred as "Model #2" (M2 hereafter) and "Model #3" (M3 hereafter). In all cases, M2 and M3 provide much better models than M1. M2 and M3 abundances of both amorphous and crystalline ice are quite stable compared to M1 abundances, except for spectra [-55°N, 270°W] and [-60°N, 80°W] where the total abundance of water ice drops strongly, down to almost 50%. A close comparison between M2 and M3 best-fit abundances reveals (i) very close results for the darkening agent and (ii) that the difference between the abundance of the sulfuric acid hydrate in M2 and the total abundances of salts in M3 never exceeds 10%. Nevertheless, some differences are observed in terms of RMS between M2 and M3. Indeed, while a slightly better fit is obtained with M2 for pixel [-60°N, 80°W], the best-fit spectrum quality is equivalent or better with M3. In general, the sulfuric acid seems useful to reproduce the shorter wavelengths, especially the 1.50 µm absorption. On the other hand, salts seem required to reproduce (i) the shoulder at 1.74 µm and (ii) the shift to shorter wavelength of the 2.02 µm absorption.

Insofar as both the sulfuric acid hydrate and salts are needed at some point to improve the fits, these two types of minerals are then used simultaneously for next set of models. However, with twelve different minerals, the salts are over-represented compared to other laboratory spectra, so a subset was used. While the carbonate – $Na_2CO_3.10(H_2O)$ – appears irrelevant in most cases, either the "Model #4" (M4 hereafter) using five different sulfates – $Na_2Mg(SO_4)_2.4(H_2O)$, $MgSO_4.6(H_2O)$, $MgSO_4.7(H_2O)$, $Na_2SO_4.10(H_2O)$ and $MgSO_4.11(H_2O)$ – or the "Model #5" (M5 hereafter) using six different chlorinated salts – $NaClO_4.2(H_2O)$, $MgCl_2.2(H_2O)$, $MgCl_2.4(H_2O)$, $MgCl_2.6(H_2O)$, $Mg(ClO_3)_2.6(H_2O)$ and $Mg(ClO_4)_2.6(H_2O)$ – generally produce better modeled spectra compared to previous models. In general, M4 and M5 are fairly equivalent, nevertheless some differences are visible for some spectra:

M4 is undoubtedly better for the pixel [15°N, 80°W], while M5 produce better results for the pixels [-55°N, 270°W] and [-60°N, 80°W]. Detailed investigations of these models show that Mg-bearing salts are generally more abundant than those Na-bearing, but this may also be an influence of the difference in hydrated states. Next, a dozen of models combining sulfates and chlorinated salts were tested. The final modeling step, named "Model 6" (M6 hereafter), uses two sulfates and two chlorinated minerals: $Na_2Mg(SO_4)_2.4(H_2O)$, $MgSO_4.6(H_2O)$, $MgCl_2.6(H_2O)$ and $Mg(ClO_3)_2.6(H_2O)$. Individually, abundances of these minerals are generally ≤ 2% and could be considered as irrelevant or negligible, but the sum of these abundances reaches 15% in some locations, suggesting that hydrated minerals are an important constituent of the surface.

A last test was performed using the whole laboratory spectral library (table 3) but no clear improvement, usually visually observed when $RMS_{TOT}$ drops by 0.0005, was observed. Thus, Ganymede's surface in SINFONI's H+K range can be properly modeled with fourteen end-members belonging to four different mineralogical families: (i) crystalline (4 end-members) and amorphous (4 end-members) $H_2O$-ice; (ii) a darkening agent; (iii) sulfuric acid hydrate; and (iv) salts, one Mg-sulfate (hexahydrate), one Na/Mg-sulfate (bloedite), one hydrated Mg-chloride and finally one hydrated Mg-chlorate.

## 2. Abundance maps

The M6 model was then applied to every pixel. As for the figure 8, all the spectra were modeled after being smoothed over 10 spectels. For consistency, input parameters $S$ and $A_i$ used to start the iteration always remain the same for the given end-members. Based on visual verifications, resulting modeled spectra are considered consistent with SINFONI measured spectra for $RMS_{TOT} \leq 0.0180$. With a mean $RMS_{TOT} = 0.0136$ and only 2% of pixels having $RMS_{TOT} \geq 0.0180$, the final calculation produces relevant modeled spectra for Ganymede's entire surface. In the following sub-sections are presented and discussed the spatial distribution of most end-members used in the models.

### 1. *The darkening agent*

To avoid a possible influence of the darkening agent's reflectance variability on the abundance of other end-members, the same spectrally dark component is used for all models. To determine the best reflectance, multiple tests were performed with reflectances from 0.15 to 0.30 because this level of reflectance is relevant for the spectra of hydrated silicates and carbon, for which only few weak non-$H_2O$ related features, precisely Si-OH bands, are expected in the H+K wavelength range, between 2.15 μm and 2.35 μm. The best mean $RMS_{TOT}$ for the whole dataset value is obtained when the reflectance of the darkening agent is equal to 0.24 (figure 9). This result is consistent with the spectrum of Callisto, close to "pure non-$H_2O$", observed by previous telescopic observations (Calvin & Clark 1991).

With a mean abundance about 52.8 ± 12.7%, the darkening agent is the dominant compound according to spectral modeling. Its abundance map showcases a strong hemispherical dichotomy (figure 10a); the darkening agent is mainly present on the trailing orbital hemisphere with maximal abundances centered slightly eastwards to the hemisphere's apex. This hemispherical dichotomy is the most visible characteristics of the abundance map, but a latitudinal gradient is also particularly observed; whatever the longitude, the maximal abundance of the dark material is always centered near the equator, between ± 25°N of latitude. Some slight distribution heterogeneities are occasionally observed at some locations, but these variations are not attributable to any geomorphological units, except craters. Excluding the craters, near the equator the abundance goes from ~50% and locally exceeds 80%. Abundances drop to ~15% for the highest latitudes, but never go below 10%. Supposing our assumptions about the presence of the dark component are correct, model uncertainties on abundances are estimated to ± 6% based on the standard deviation of multiple squares of 5 × 5 pixels, corresponding to the effective (PSF limited) spatial resolution of the VLT. In addition, a variation of ± 0.03 in the reflectance level of the darkening agent (figure 9) engenders a mean variation of about ± 5%. Hence, final error bars on the abundance of the darkening agent is assumed to be the quadratic sum of these errors, about 8%.

2. <u>$H_2O$-ice</u>

The total $H_2O$-ice abundance map (figure 10b), composed of the sum of all grain sizes of both amorphous and crystalline $H_2O$-ices, is well anti-correlated to the dark component's abundance map (figure 10a). The highest abundance almost reaches 60% at Osiris crater while the lowest is about 10% close to the trailing hemisphere apex. A strong, but asymmetric, latitudinal gradient is observed: the higher the latitude, the higher the abundance. Overall, the distribution of $H_2O$-ice better matches that of the $BD(1.65)$ map than the $BD(2.02)$ one (figure 4), suggesting that the $H_2O$-ice is predominantly crystalline. However, while Galileo Regio and the less-cratered part of Perrine Regio are identifiable in the $BD(1.65)$ map, these regions are not identifiable any more with the abundance map. Only large and bright craters, such as Tros, Cisti and Osiris, are observed. Bright areas, such as the northern part of Xibalba sulcus and the southern part of the leading hemisphere, are enriched in $H_2O$-ice. A detailed investigation, distinguishing amorphous versus crystalline $H_2O$-ices and large versus small ice grains, is given below to better understand the overall distribution of $H_2O$-ice and the processes that create it.

i. *Crystalline versus amorphous ice*

Qualitatively, crystalline and amorphous ice have very different spatial distributions: while the crystalline ice abundance map (figure 10c) globally follows the same distribution trend as the abundance map of total $H_2O$-ice (figure 10b), the amorphous ice appears to be distributed quite homogeneously over the surface (figure 10d), with a light latitudinal gradient and weak ice depletions close to the trailing hemisphere apex and at the center of Galileo Regio. This differing distribution between crystalline and amorphous ice is likely the result of different processes that will be discussed further in the paper. Quantitatively, crystalline ice is approximately 2.5 times more abundant than amorphous ice: over the whole investigated surface, the mean abundance for the crystalline ice is 20.7 ± 7.7%, while the mean abundance for the amorphous ice only reaches 8.5 ± 3.3%. Such dominance of the crystalline phase is relatively surprising considering the results of Hansen & McCord (2004) who determined, based on the analysis of the 3.1 μm absorption band of the $H_2O$-ice, that amorphous and crystalline phases are present in equivalent proportions. It suggests the existence of a positive vertical crystallinity gradient in the ice slab, as advocated by Hansen & McCord (2004) in their paper. This vertical gradient is mainly a consequence of the range and energy loss of the charged particles in the ice slab: as the penetration range decreases with the mass of the particles the amorphization of the $H_2O$-ice grains on the surface should be dominated by lower (keV) energy protons and heavy ions (Famá et al. 2010). Overall, except for very few pixels (5% of the map) mainly located on the trailing hemisphere, crystalline ice is dominant on Ganymede's surface. The crystalline ice is mostly concentrated on the leading hemisphere. Craters are enriched in crystalline ice, where the abundance almost reaches 50% at Osiris crater. The abundance drops below 5% at the equator of the trailing hemisphere, inside Melotte Regio. For amorphous ice, the it goes up to 22% at high latitudes and is close to 0% at Galileo Regio and at trailing hemisphere's apex.

### ii. *The grain size spatial distribution*

There are strong differences in terms of spatial distribution between the different grain sizes of the $H_2O$-ice. The small grain abundance map (figure 10e), grouping 10 μm and 50 μm grains of crystalline and amorphous ice, shows a very strong latitudinal gradient in contrast with the large grain abundance map (figure 10f), composed of 1 mm grains of crystalline and amorphous ice. The small grain map showcases correlations with geomorphological units: fresh, bright, craters (such as Osiris, Cisti, Tros) and sulci look enriched while most of the regiones (Galileo, Melotte, the southern part of Perrine and the observed part of Bernard) are depleted. Quantitatively, in the latitudes ± 30°N the mean abundance of smaller grains is about 6.2 ± 2.0% after excluding fresh, bright, craters. Then, there is a sharp transition around ± 35°N beyond which the abundance increases very rapidly, reaching 45% at Osiris crater and in the northern part of Xibalba sulcus. Conversely, the equator and mid-latitudes are mostly dominated by the large grains. Correlations with geomorphological units are identifiable, such as Galileo Regio hosting the highest abundance for large grains in its center, slightly exceeding 25%.

### 3. *Sulfuric acid (octa-)hydrate*

Quantitatively, the mean abundance of the sulfuric acid hydrate all over Ganymede's surface (figure 10g) is 9.6 ± 5.5%. The range of abundances starts from ~0% at the equator on the trailing hemisphere and almost reaches 30% in the south of the leading hemisphere. The sulfuric acid hydrate end-member of the spectral modeling is mostly used for pixels in the leading orbital hemisphere and at high latitudes, contrary to what was initially expected based on its distribution over the surface of Europa (Carlson et al. 2005, Ligier et al. 2016). Overall, the spatial distribution is uneven and not really correlated with geomorphology, but looks very well anti-correlated with the darkening agent (figure 10a). The question about the spatial distribution of this end-member will be addressed in the discussion.

### 4. *Salty minerals*

The mean abundance of the total salts map (figure 10h) is 8.5 ± 3.0%, similar to amorphous ice. Significant concentrations, reaching up to 19%, are found at different areas: (i) the easternmost part of Harpagia Sulcus [-12°N, 314°W], (ii) high latitudes of the trailing hemisphere and (iii) the surroundings of the southern part of Perrine Regio, which is separated from others regions by multiple sulci: Xibalba at west, Babylon [-23°N, 96°W] at south, Nineveh [23°N, 53°W] at north and finally Phrygia [12°N, 21°W] and Sicyon [34°N, 15°W] at east. The lowest abundances are located in Marius Regio, close to the anti-Jovian point [0°N, 180°W], and vary between 1% and 7%. However, following the same method used to estimate the abundance uncertainties of the darkening agent, in the worst cases uncertainties about 5% for salts are obtained. These uncertainties seem representative of the detection threshold of the salts; thus, pixels with abundances lower than 5% are considered "salt-free". Qualitatively, the salt total abundance map does not show clear correlation with any other maps previously presented: neither the hemispherical dichotomy nor a clear latitudinal gradient are observed. A detailed investigation distinguishing the sulfates from the chlorines is necessary to better the separation of the salts. Such an investigation may also bring new insights about the origin and eventually the processes engendering the formation of these minerals.

### i. Sulfates

Among the non-$H_2O$ minerals, sulfates are those with the most localized, restricted, spatial distribution (figure 10i). These salts are almost exclusively located in sulci: Harpagia, Xibalba (its southernmost part), Babylon (its northernmost part), Phrygia, Sicyon and Nineveh. The abundance in these sulci is around 8% in average, and the maximal abundance goes up to 14% for some pixels in Harpagia, Xibalba and Babylon. Supplementary tests were performed to confirm this peculiar spatial distribution, especially by varying the reflectance level of the darkening agent: with a reflectance level varying from 0.21 to 0.27 it appears that (i) the darkening agent does not influence this distribution and (ii) the abundances remain similar at a ± 3% level in average. Furthermore, the standard deviation in a 5 × 5 pixels square in these enriched sulci does not exceed 3% either, thus bringing confidence over the presence of sulfates there. After excluding the enriched sulci, the mean abundance over the entire surface drops below 3%, i.e. considered as "sulfate-free". Nevertheless, as demonstrated previously in this section when describing the selection procedure of the end-members, their presence is needed to improve the quality of the fit, especially for the spectrum (15°N, 80°W) shown in figure 8, corresponding to the south of Xibalba sulcus and for which the total abundance of sulfates is 9.5%.

### ii. Chlorinated salts

Like sulfates, maximum abundances of chlorinated salts are about 14% (figure 10j), but enriched areas are located elsewhere, specifically in the trailing hemisphere and especially south to the Tashmetum crater. Lowest abundances are near the equator and are centered on the anti-Jovian point. Like sulfates, the chlorinated salts have a heterogeneous spatial distribution, however they are more dispersed on Ganymede's surface and their distribution does not seem correlated to known geological units. Overall, chlorinated salts as modeled appear slightly more abundant than sulfates (4.8 ± 2.1% for chlorines vs 3.7 ± 2.4% for sulfates). Similarly to sulfates, despite the fact that the derived abundances for the chlorinated salts remain very low compared to the ice and "dark" materials, their inclusion in the spectral model does significantly improve the fits of some spectra, as shown in figure 8 and table 4. The origin of these salts is discussed in the next section.

**DISCUSSION**

1. **Surface roughness**

As mentioned and explained previously (figure 3), Oren-Nayar's roughness parameter, $\sigma$, is determined empirically by performing multiple tests until a value provides a fit to Ganymede's albedo pattern getting few, or no, weak high inclinations residuals. This approach is quite different than the more common approach using a radiative transfer model, in most cases Hapke's model (Hapke 1981, Hapke 1984), to derive many parameters including the degree of macroscopic surface roughness, respectively called $\theta$ or $\theta$-bar (e.g. Buratti 1991, Domingue & Verbiscer 1997). Here, roughness values from $\sigma = 16° \pm 6°$ on the trailing hemisphere to $\sigma = 21° \pm 6°$ on the leading hemisphere are obtained. The values derived from Hapke's radiative transfer model: (i) $\theta = 29° \pm 2°$ on the trailing hemisphere for Buratti (1991), and (ii) $\theta$-bar $= 28° \pm 5°$ or $\theta$-bar $= 35° \pm 5°$ respectively on the leading or the trailing hemisphere for Domingue & Verbiscer (1997). Qualitatively, like those of Domingue & Verbiscer (1997), the results of this study may show a subtle hemispherical dichotomy for the roughness, but this dichotomy is found to have an opposite distribution than that found in the previously published studies; here, it is the leading hemisphere that gets the highest $\sigma$ values. This possible dichotomy could be explained by enhanced meteorite and micrometeorite bombardment of the leading hemisphere. This hypothesis is supported by the density of bright, fresh, impact craters observed on the leading orbital hemisphere compared to the trailing one (figure 4). Nevertheless, this result has to be taken cautiously due to the large uncertainties in our estimates of $\sigma$. Lastly, one has to keep in mind that the differences between the results presented here and those obtained previously might be explained by the use of different model. Oren-Nayar's model remains simplistic compared to Hapke's model, but it has been demonstrated that it is still an undeniable improvement compared to the Lambertian model and that it results in a very solid photometric correction up to $\theta_i = 65°$ (figure 2, figure 3).

2. **Sulfuric acid hydrate**

The sulfuric acid hydrate is one product of radiolysis of surface ice by high energy particles from the Jovian magnetosphere, such as $S^{n+}$ and $O^{n+}$ cations (Carlson et al. 1999, Johnson et al. 2004, Strazzulla et al. 2009, Ding et al. 2013). These charged particles, very dominant in the Jovian magnetosphere due to the intense volcanic activity from Io (Thomas et al. 2004), are trapped within Jupiter's rotating magnetosphere and can reach Ganymede's surface. At Ganymede's orbit, Jupiter's plasma is close to co-rotating with Jupiter (Williams et al. 1997). Because Ganymede synchronously rotates as it revolves around Jupiter about 17 times slower than the corotational speed of the plasma, Ganymede is constantly overtaken in its orbit by plasma and energetic particles trapped in Jupiter's magnetosphere, and the trailing hemisphere is always facing the magnetospheric plasma flow. Therefore, Ganymede's trailing hemisphere should be enriched in S-bearing minerals produced by radiolysis, such as sulfuric acid hydrate. This hypothesis has already been tested and verified for Europa (Carlson et al. 2005, Grundy et al. 2007, Brown & Hand 2013, Ligier et al. 2016). However, since Ganymede's intrinsic magnetic field was discovered by the Galileo mission (Kivelson et al. 1996), it is expected that the Jovian magnetospheric plasma flow does not have direct access onto equatorial latitudes of Ganymede between the north and south open/closed field lines boundaries. Since that discovery, studies have been undertaken to model Ganymede's magnetic environment and constrain its influence over the surface of the satellite (Khurana et al. 2007, Allioux et al. 2013, Plainaki et al. 2015, Fatemi et al. 2016, Poppe et al. 2018). All these models do highlight that the cold and hot ion plasma flux onto Ganymede should be much lower at lower latitudes (especially on the trailing hemisphere), i.e. between ± 35°N of latitude. Thus, the poles should be more heavily weathered. In addition, the most recent studies predict a hemispherical asymmetry where the flux of incident charged particles is higher on the leading hemisphere (Plainaki et al. 2015, Fatemi et al. 2016, Poppe et al. 2018). All these results are consistent with the abundance map of the sulfuric acid hydrate presented here (figure 10g). Indeed, this map exhibits that (i) most of the sulfuric acid is located at high latitude and (ii) mainly on the leading hemisphere. Hence, despite a spatial distribution opposite to the one on Europa (Ligier et al. 2016), the sulfuric acid hydrate on Ganymede's surface is very likely to be present because of exogenous inputs (namely sulfur ions). But, in accordance with Fatemi's (2016) and Poppe's (2018) models of total precipitating ion flux, northern high latitudes of the trailing hemisphere should be slightly richer in sulfuric acid hydrate compared to what is obtained for figure 10. However, these slight discrepancies are a matter of few percent in terms of abundance and might be explained by changing grain sizes for the sulfuric acid hydrate end-member used in the modeling.

3. **$H_2O$-ice**

To first order, the abundance maps of $H_2O$-ice are dominated by a latitudinal gradient. Then, at second order comes spatial correlations with geomorphological units. The latitudinal gradient is generally linked to the impact of the external environment, while geomorphological correspondences may be potentially related to geological, i.e. endogenous, processes. The spatial coverage and resolution of our dataset brings important constraints on the balance between these two different processes. For reading convenience, the influence of the external environment is discussed in the first sub-section, and questions about the spatial correlations with some geomorphological units is addressed in the second.

*1. Influence of the external environment*

Figure 10b shows that $H_2O$-ice is nearly absent around the apex of the trailing hemisphere and tends to be less abundant near the equator. On the Saturnian satellites, it has been argued that impacting grains can "sandblast" the ice, causing fresh, brighter ice to be observed where grain impacts are high (Verbiscer et al. 2007). On Ganymede, because of its intrinsic magnetic field, the high latitude regions are more heavily sputtered and the leading is more sputtered than the trailing along the equator (Plainaki et al. 2015, Fatemi et al. 2016, Poppe et al. 2018). One should note that these simulations generally corresponds to a situation where Ganymede is located at the center of the Jovian plasma sheet. When Ganymede is at other Jovian magnetic latitudes, it is possible that the precipitation pattern would shift slightly. In any cases, these models are consistent with the observations, i.e. brighter ice in the poles and along the leading equator. However, complicating this story is the idea that the equatorial regions are also where sublimation is the highest. Plainaki et al. (2015) predict that this process is the most important one occurring on Ganymede for water molecule release. Combined with polar temperatures, this would preferentially redistribute ice from the equator to the colder poles, observed in figure 10b. But, if the sublimation process dominates as proposed by Plainaki et al. (2015), this does not explain why the trailing equatorial region would be more depleted in $H_2O$-ice than the leading equatorial region. Other exogenous processes are therefore envisaged as the cause of the hemispherical dichotomy at the equator; the most evident process is the surface brightening due to ice excavation, i.e., the removal of the upper layer weathered ice and the exposure of fresh ice below, engendered by the meteoritic and micrometeoritic bombardment. This bombardment, which preferentially occurs on the leading orbital hemisphere (Bottke et al. 2013), may become the main exogenic process to explain the distribution of $H_2O$-ice over the surface shielded by Ganymede's own intrinsic magnetic field.

Then, there is the question of whether this picture fits together with the global distribution obtained for both amorphous and crystalline ice (figures 10c and 10d). Figure 10c shows the crystalline ice, which tracks the distribution of water ice generally (figure 10b). Figure 10d shows the amorphous ice which should mostly be located in the very first micrometer of the surface. Indeed, at least one study predicted the amorphous ice was due to higher radiation processing of the surface water ice (Hansen and McCord 2004). Laboratory experiments have proved that when water ice is bombarded with energetic ions at cold temperatures, it can lead to the amorphization of the ice (Famá et al. 2010). This only occurs if the timescales for that process are faster than the timescales of thermal recrystallization, assuming other variables are the same. To understand whether this is a dominant process, we consider the most abundant energetic ions hitting the surface. The dominant charged particle in the 100 keV to few MeV range near Ganymede is protons (Mauk et al. 2004). Allioux et al. (2013) find that energetic proton access patterns are similar to cold plasma ones, i.e. stronger at polar regions. Figure 10d does indeed show weakly higher abundance of amorphous ice within the polar regions. Last but not least, we note the competing processes is also very important: at high latitude, the thermal recrystallization's rate is slower than at the equator due to the colder temperatures.

To conclude, the grain size distribution seems correlated with the Jovian magnetospheric bombardment, at least with the distribution of the smaller grains (figure 10e), showing, at first order, a strong latitudinal variation with a quite sharp transition around ± 35°N for the northern hemisphere (figure 11). First of all, these maps are in agreement in terms of trend and in terms of grain size with former spectral analysis of Galileo/NIMS data (Hibbits et al. 2003, Hansen & McCord 2004). In addition, the distribution seems to follow very nicely the ion flux maps modeled by Fatemi et al. (2016) and Poppe et al. (2018). Nevertheless, we note the smallest grains are found in the region of highest predicted sputtering (Fatemi et al. 2016). Previous work has suggested that sputtering would tend to erode faster smaller grains, thus leading to a larger average grain size (Clark et al. 1983). However, this process does not appear to be dominant on Ganymede in shaping the final grain size. Actually, at Ganymede's orbit, it has been proposed that the sputtering of ice by energetic particles would lead to local redeposition of finer grained ice (Johnson 1997, Khurana et al. 2007). In addition, the crystal growth is exponentially dependent on temperature (Clark et al. 1983). Hence, the combining effect of these two processes may inhibit the crystal growth and could do better than simply compete against it. To conclude on the grain size topic, it is worth mentioning that Johnson and Jesser (1997) predict that energetic electron bombardment may result in the production of small voids in grains, leading to increasing internal light scattering and thus artificially diminishing the grain size needed in our spectral modeling. If this process is a dominant one, it would be another complication to grain size determination.

### 2. *Geomorphological correlations*

After first order spatial correlations because of exogenous processes, second order spatial correlations are related to geomorphological units highlighted on the USGS map (figure 4). The spatial correlations are mostly identified on maps related to crystalline ice (figure 10c) and mainly associated with small grain sizes (10e and 10f): they are either related to craters (especially the brighter in the visible, such as Osiris, Cisti and Tros) and large sulci (Xibalba, Babylon) which are enriched in ice; conversely darker regiones are ice-depleted. On one hand, the excavation of fresh ice from Ganymede's icy crust caused by impact craters is then followed by a local redeposition of this excavated ice. On the other hand, the localized enrichment or depletion of small or coarse grains, as observed in some sulci and regiones, is intriguing and has never been observed so far. Such abnormal spatial distribution is generally considered as the footprints of recent endogenic activity. More precisely, freezing liquid water flows would create large grains while finely grained ice particles could be produced with the activity of cryogenic plumes (Clark et al. 1983). Unfortunately, evidence of cryovolcanic activity is not supported by Galileo high resolution images (Pappalardo et al. 2004). Moreover, based on surface cratering rates, regiones and sulci are respectively aged about >4 Gyr and 2 Gyr in average, and possibly around 0.4 Gyr for the youngest sulci (Zahnle et al. 1998, Pappalardo et al. 2004). Since the emplacement, the annealing and the sintering of ice particles should have implied a sort of spatial homogenization in terms of grain size, which is not observed here (figure 10e, 10f). The direct responsibility of endogenic processes on the grain size distribution seems quite unlikely, or at least secondary. But the different albedo between regiones and sulci could lead to local differences in terms of surface temperature, which affects the grain growth (Wilson 1982). It is therefore expected that the dark (hotter) terrains, e.g.

Galileo Regio, promote more effectively the grain growth than the bright (colder) terrains like sulci and fresh craters (Khurana et al. 2007).

## 4. Darkening agent

The spectral modeling of our data confirms different hypotheses proposed in former studies: (i) it confirms that the darkening agent is the main component of Ganymede's surface in equatorial and mid-latitudes, (ii) it also confirms the results of Calvin & Clark (1991) and Helfenstein et al. (1997), obtained respectively from ground-based observations and Galileo images, who found a reflectance level for the non-$H_2O$ material between 0.20 and 0.30 in the H+K wavelength range, and (iii) it finally confirms that there are some areas with ≥80% of non-$H_2O$, dark material (Pappalardo et al. 2004). In addition to these results, the global coverage of Ganymede's surface by our ground-based campaign bring constraints on the spatial distribution of this darkening agent (figure 10a), which is clearly related to the Jovian expected bombardment pattern of magnetospheric ions. As expected, the distribution is well anti-correlated to both amorphous and crystalline water ice and/or sulfuric acid (figure 12). The simplest solution to explain such anti-correlation consists of looking at the main consequence of the surface sputtering: the stronger the sputtering, the more eroded the surface and the more particles are ejected (Johnson et al. 2004). These particles will then be redistributed across Ganymede's surface, mostly locally (Johnson 1997, Khurana et al. 2007). We hypothesize here that the surface sputtering and micrometeoroid gardening lead to the deposition of fresh ice particles that eventually cover the darkening agent, thus explaining the distribution observed figure 10a. Finally, in terms of composition, the nature of the darkening agent is still unidentified. Indeed, even at SINFONI's spectral resolution which is fifty times better than NIMS', no features due to non-water-ice-related material has been observed in the spectra. Such absence of sharp absorptions features, combined with the reflectance level required, lead to consider once again hydrated silicates as excellent spectral candidates for the darkening agent (Calvin & Clark 1991).

## 5. Salty minerals

The spectral modeling showcases the need for hydrated salty minerals to better fit SINFONI's measurements (figure 8). Two types of salts are required: chlorinated and sulfates. Unlike the total $H_2O$-ice, the sulfuric acid hydrate and the darkening agent, the spatial distribution of salts is neither related to the Jovian magnetospheric environment nor the craters (figure 10h). As mentioned earlier, the salts are distributed very locally, especially sulfates (figure 10i); these areas correspond to sulci, which are the youngest areas of Ganymede's surface (Zahnle et al. 1998, Pappalardo et al. 2004). Furthermore, as predicted by multiple meteoritic leaching experiments, ions ($Na^+$, $Mg^{2+}$, $SO_4^{2-}$, $Cl^-$, etc…) are leached from Ganymede's silicate interior because of water circulation and interactions of these ions with liquid water form hydrated salts that are eventually emplaced on the surface thanks to upwelling of material through the ice shell (Kargel 1991, Kargel et al. 2000, Fanale et al. 2001, Zolotov & Shock 2001). Consequently, the presence on the surface of these minerals is very likely due to geological processes. The existence of Mg-sulfates and/or Na-sulfates is long expected on the Galilean icy moons insofar as $Mg^{2+}$, $Na^+$ and $SO_4^{2-}$ ions are found in numerous studies to be by far the most abundant ions in the ocean (Kargel et al. 2000, Fanale et al. 2001, Zolotov & Shock 2001). Concerning the chlorinated salts, their existence has long been neglected due to the much weaker expected concentration of the $Cl^-$ ion in Ganymede's ocean, estimated to be ten times lower in abundance than $SO_4^{2-}$ by Kargel et al. (2000). However, fractional crystallization of the upwelling fluid would increase rapidly the $Cl^-/SO_4^{2-}$ ratio, thus allowing the emplacement of chlorinated salts on the surface of the Galilean icy moons (Kargel et al. 2000, Zolotov & Shock 2001). Nevertheless, the presence of both sulfates and chlorinated salts used in the modeling cannot be definitely asserted. First, because their spectral effects on the model are very subtle even at SINFONI's spectral resolution (figure 8). Second, because they share some spectral similarities with other salts (Dalton et al. 2005, Dalton & Pitman 2012, Hanley et al. 2014). Finally, linear unmixing remains a first-order approach and does not take into account the influence of changing grain size. Hence, two major steps have to be performed in order to confirm or disprove the presence of these minerals: (i) new experiments to get the optical constants of the minerals, thus allowing a nonlinear unmixing approach, and (ii) acquisition of new measurements combining very high spatial and spectral resolutions, i.e. data obtained from an instrument on an orbiter (like MAJIS on JUICE or MISE on the NASA's Europa Clipper mission).

## 6. Residual absoprtions

Overall, the modeled spectra reproduce quite well the measured spectra provided in figure 8, even if some small differences are still observed. To check the global consistency of the modeling and to detect possible spectral residues, the $data/model$ ratio is calculated using the whole dataset (figure 13). Firstly, this figure shows that the $data/model$ ratio over the range is mostly between 0.985 and 1.015, which is larger than the noise. But, similarly to the figure 8, it also highlights three main discrepancies: (i) in the 1.46 – 1.55 µm range, (ii) in the 1.94 – 1.97 µm range and finally (iii) around 2.24 µm. The two first ones are related to the main $H_2O$-ice absorption bands, but with opposite trends, while the last one shows a hollow pattern that might be characteristics of a spectral absorption from a missing compound. In the 1.46 – 1.55 µm range, the ratio decreases slowly and progressively with increasing wavelength: this is quite typical of a grain size effect, likely attributed to the hydrated salts and/or the sulfuric acid hydrate insofar as varying the grain sizes of $H_2O$-ice does not produce clear improvements in this range. In the 1.94 – 1.97 µm range, the ratio decreases towards shorter wavelengths. The distortion of the 2.02 µm $H_2O$-ice absorption band is also on observed on Europa and is usually attributed to the existence of unidentified heavily hydrated minerals (McCord et al. 1998b, Carlson et al. 1999, Brown & Hand 2013, Ligier et al. 2016). Finally, the residue at 2.24 µm is probably the most intriguing feature. Its position is uncorrelated to any H-O-H or O-H or molecular vibration, so it should be characteristics of a non-$H_2O$ material with a wide absorption.

To further investigate the residue at 2.24 µm, its related band depth map has been created (figure 14). The position of the band center and its left/right shoulders is shown in figure 13: 2.120 µm for the left shoulder, 2.240 µm for the center and 2.315 µm for the right shoulder. Like the $BD(2.02)$ and $BD(1.65)$ maps (figure 4), the values used to obtain figure 14 correspond to the median reflectance value over a ± 0.0025 µm-range (i.e. 11 wavelengths). It is very interesting to note that, in addition to being distributed heterogeneously over Ganymede's surface, this map shares numerous similarities with the abundance map of the chlorinated salts (figure 10j). Based on this observation, one can suggest two hypotheses: (i) at least one chlorinated end-member with a wide absorption around 2.24 µm is missing in the spectral modeling, (ii) the chlorinated salts used in the modeling are irrelevant and could be replaced by one or more minerals relatively flat in the H+K wavelength domain but still with a wide absorption around 2.24 µm. In our opinion, both hypotheses are equally likely. Indeed, about the first hypothesis, $MgCl_2$ and NaCl cryogenic brines do possess an absorption there, but it might not be wide enough and it is centered at 2.22 µm instead of 2.24 µm (Hanley et al. 2014, Thomas et al. 2017). Concerning the second hypothesis, hydrated silica-bearing minerals, such as opal and chalcedony, exhibit a wide absorption band around 2.24 µm that could fit the residue (Langer & Flörke 1974, Goryniuk et al. 2004). In addition, these minerals also have a strong absorption at 1.90 µm that should help to improve the fit in the 1.94 – 1.97 µm range. Last but not least, these minerals are generally associated with hydrothermal systems (Goryniuk et al. 2004), which is particularly interesting in terms of endogenous processes and implications for extraterrestrial life. Nevertheless, none of these hypotheses can be asserted with this dataset. Much better spatial resolution, e.g. with the MAJIS of the JUICE mission, seems absolutely required to do so.

## CONCLUSION

The high spectral and spatial resolution data acquired with SINFONI allowed us to revisit the properties and composition of Ganymede's surface. First, the photometric corrections highlight the fact that the surface of Ganymede does not have a Lambertian photometric behavior, unlike Europa (Ligier et al. 2016). This initial result is in agreement with a globally much rougher surface at Ganymede than Europa. Then, at SINFONI's spectral resolution, we have concluded that the spectral properties of Ganymede cannot be reproduced by a model using only crystalline and amorphous $H_2O$-ice and a spectrally flat darkening agent: sulfuric acid hydrate and salts (sulfates and chlorinated) seem needed to well fit the observed spectra.

Concerning the $H_2O$-ice, its amorphous and crystalline phases are both present on the surface, but the crystalline is almost 2.5 times more abundant (overall, 20.7 ± 7.7% for the crystalline versus 8.5 ± 3.3% for the amorphous). Overall, the spatial distribution of $H_2O$-ice, and especially its crystalline phase, is very clearly dominated by a latitudinal gradient. It is likely consistent with sublimation and preferred condensation near the poles. The leading/trailing asymmetry may be explained by a slightly higher sputtering rate and/or preferential micrometeoroid impacts on the leading hemisphere. The fraction of crystalline water ice generally tracks the total water ice abundance. The amorphous ice is more abundant in polar regions, which are much more heavily weathered by magnetospheric ions. The global scale modeling confirms that Ganymede's surface at high latitudes (beyond ± 35°N) is almost entirely covered by small grains (i.e. ≤ 50 μm), while mid-latitudes and equatorial regions are globally covered by larger grains (i.e. ≥ 200 μm, up to 1 mm). A modest hemispherical dichotomy is also observed for the smaller grains, where the leading hemisphere looks enriched. Once again, it may be explained by the leading hemisphere's stronger Jovian ion flux, which leads to higher sputtering rates and consequently to smaller grains, and/or by the leading hemisphere's higher impact cratering rate, that does show numerous fresh craters. Lastly, spatial correlations with some geomorphological units (especially craters) are observed for small and large grains. Excavation of fresher ice can be involved for the craters, while a differential grain growth rate due to local differences in terms of albedo and/or temperature was active for other regions.

The analysis shows that the best darkening agent for the modeling corresponds to a flat spectrum within the SINFONI H+K range with a high reflectance level, here 0.24, thus confirming results of former studies (Calvin & Clark 1991, Calvin et al. 1995, Helfenstein et al. 1997). The overall distribution is well anti-correlated to the $H_2O$-ice, but also exhibits a hemispherical dichotomy: the trailing hemisphere is enriched by slightly more than 10% (the abundance exceeding 80% near the apex) compared to the leading hemisphere. The covering of the dark compound by freshly crystallized $H_2O$-ice particles, originally eroded from Ganymede's surface through magnetospheric sputtering and micrometeoroid gardening, is certainly the simplest, and the best, solution to explain the distribution.

Compared to the $H_2O$-ice and the darkening agent, the sulfuric acid hydrate and the salts are secondary minerals as they respectively represent on average less than 10% of the surface composition, but they seem to be essential to model most of the spectra. Their distribution across the surface is different: the sulfuric acid hydrate is correlated to $H_2O$-ice (therefore anti-correlated to the darkening agent), while the distribution of the salts is neither related to the expected bombardment patterns of magnetospheric ions or crater locations. For the sulfuric acid hydrate, the distribution is consistent with the expected precipitation patterns of magnetospheric ions, thus confirming its formation by radiolysis of $H_2O$-ice and implantation of S ions. For the salts, an endogenous origin is highly suspected because they are mostly located on sulci. The partial fractional crystallization of upwelling fluids moving towards the surface may explain the presence of both sulfates and chlorinated salts on the surface. Last but not least, spectral residue probably related to chlorinated brines or hydrated silica-bearing minerals is observed where the chlorinated salts are located. This result could have very important implications in the search for extraterrestrial life.

Future perspectives include the use of a nonlinear (e.g. radiative transfer) approach for the spectral modeling, but optical constants for the relevant spectral end-members at cryogenic temperature is needed first. Extending the analysis and modeling across a wider wavelength range, especially beyond 3.5 μm where organic absorptions are much stronger, will also be worthwhile. Finally, increasing the spatial resolution to few kilometers will certainly be useful, to have a better view of local variation in terms of composition. Additional small spectral absorptions may exist at kilometer scales. The MAJIS instrument of the upcoming JUICE mission and the instruments of the Europa Clipper mission will hopefully address this question.

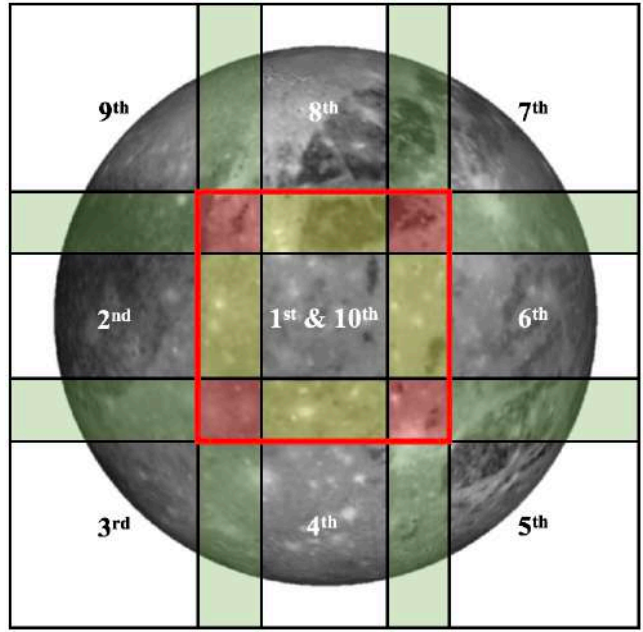

**Figure 1**. Typical observation mosaic. Central frames (1$^{st}$ and 10$^{th}$, edges in red) of the mosaic overlap the eight edge pieces via, at least, one of the four small squares coloured in red. Pixels in these small red squares and in the four yellow rectangles are used to normalize each frame with the first one. Additionally, the pixels in the green rectangles belong to different edge frames and are used to normalize adjacent frames. After the normalization, all these coloured pixels are weighted then averaged to smooth differences that might exist between two neighbouring frames. Final result consists in one unique 3D-cube for each night of observation.

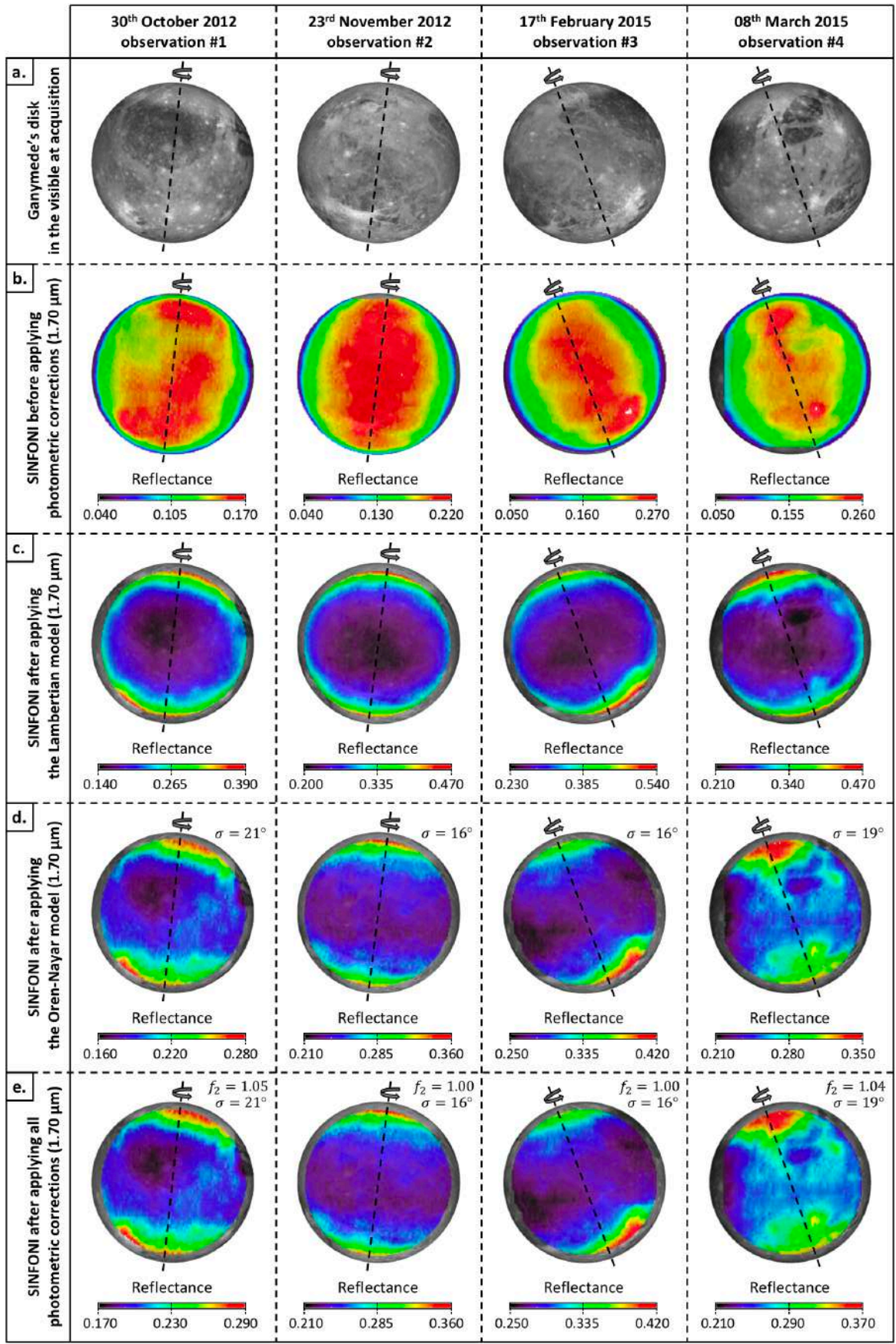

**Figure 2**. Photometric corrections, step by step. Ganymede's disks at acquisition have been obtained from the following site: http://aa.usno.navy.mil/data/docs/diskmap.php. Photometric corrections are applied for $\theta_i \leq 65°$; beyond that angle, the Oren-Nayar model is no longer valid insofar as pixels are over-corrected, similar to pixels with $\theta_i \geq 30°$ in the Lambertian model.

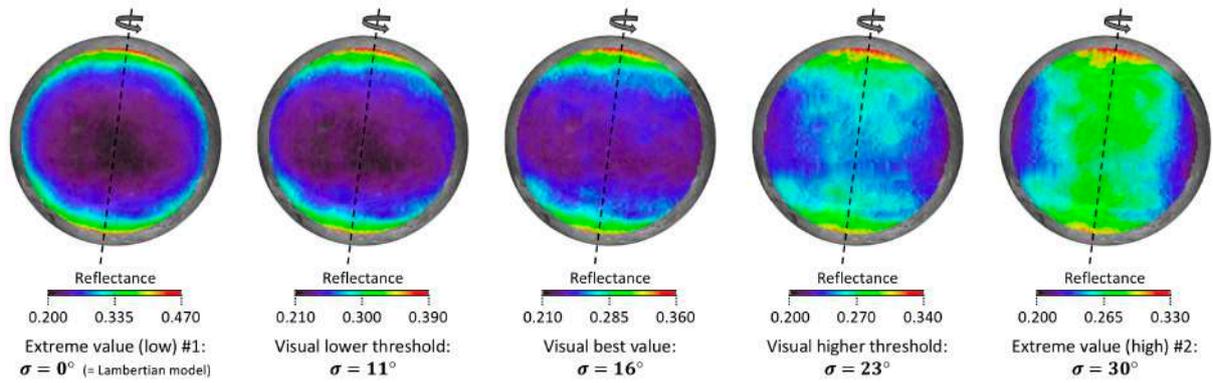

**Figure 3**. Example of the visual determination of the roughness ($\sigma$), here for the 23/11/2012 observation at 1.70 μm. If $\sigma$ is under-estimated ($\sigma \leq 11°$ in this case), the reflectance at the edge of Ganymede's disk is over-estimated compared to the center, i.e. the sub-solar point. More precisely, the global trend is that pixel's reflectance increases with the distance from the sub-solar point. On the other hand, if $\sigma$ is over-estimated ($\sigma \geq 23°$ in this case), the reflectance of the disk center is over-estimated compared to high inclination pixels. $\sigma$'s higher threshold is more difficult to determine than $\sigma$'s lower threshold because of poles' high reflectance in the infrared. Furthermore, spatial correlation with geomorphological units makes this determination not straightforward.

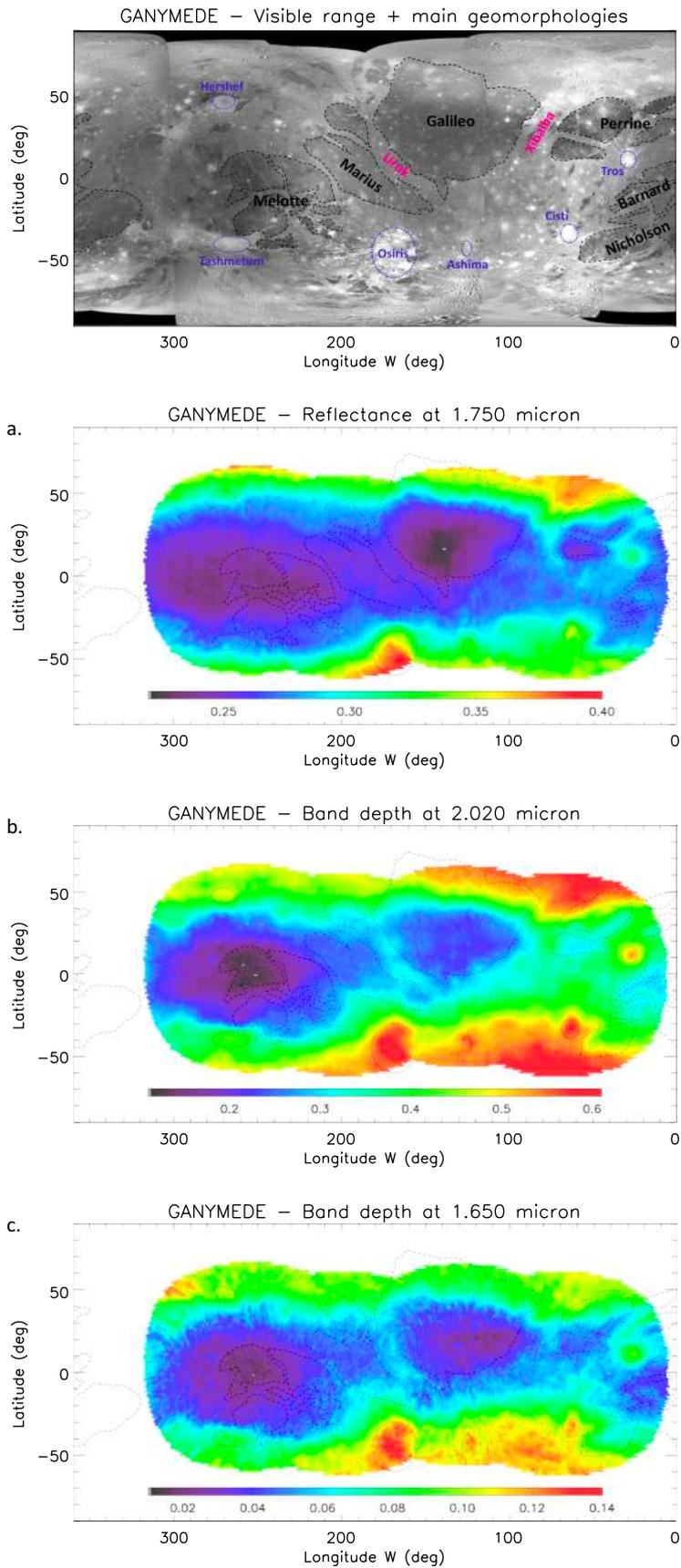

**Figure 4**. Ganymede USGS map in the visible compared to three maps: from top to bottom, (a.) the reflectance at 1.75 μm, (b.) $BD(2.02)$ and (c.) $BD(1.65)$. Black dashed lines delimit the Regio and violet dotted lines delimit the craters. Good spatial correlations with some geological units are observed whatever the map. The maps are displayed in equirectangular projection.

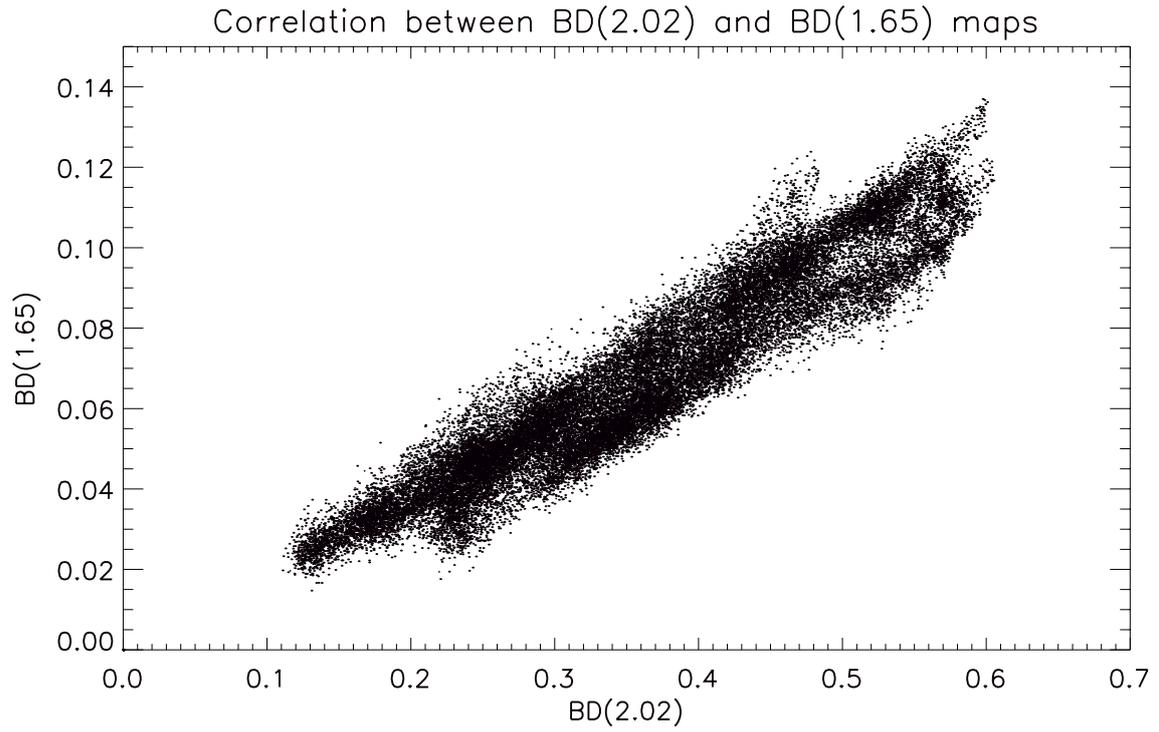

**Figure 5**. Scatter plot showing $BD(1.65)$ as a function of $BD(2.02)$. Each dot represents a single pixel of the band depth maps (Figure 4b,c). A global trend is observed, highlighting the good correlation between these maps.

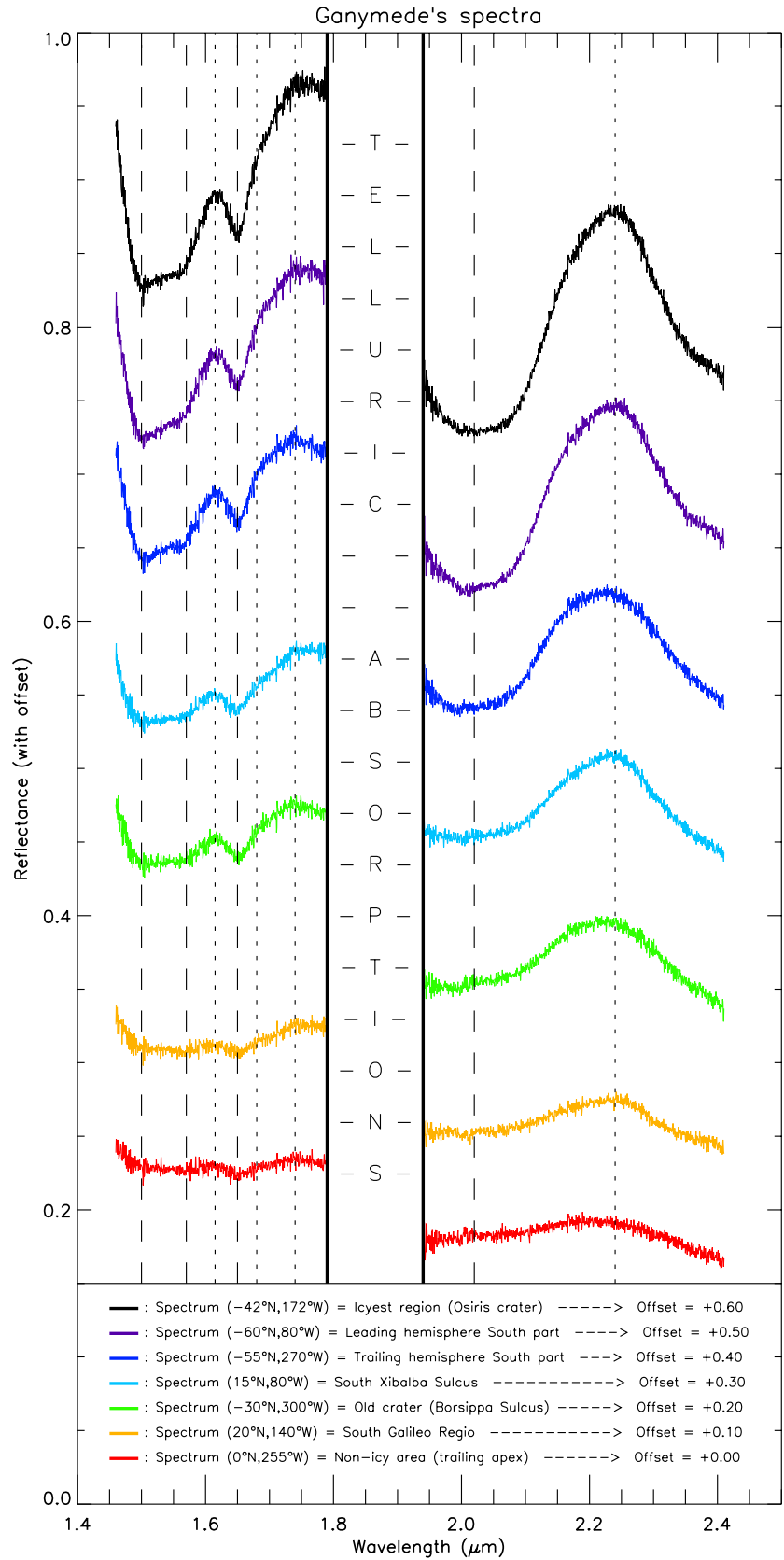

**Figure 6**. Different single-pixel spectra of Ganymede's surface, from the less icy (bottom) to the icyest (top) based on band depth maps (Figure 3). Dashed lines locate $H_2O$-ice absorptions bands: 1.50 μm, 1.57 μm, 1.65 μm and 2.02 μm. Dotted lines highlight the left and right bounds of the two latter ones, used to get $BD(2.020)$ and $BD(1.650)$. Telluric absorptions, from 1.79 μm to 1.94 μm, have been removed for clarity.

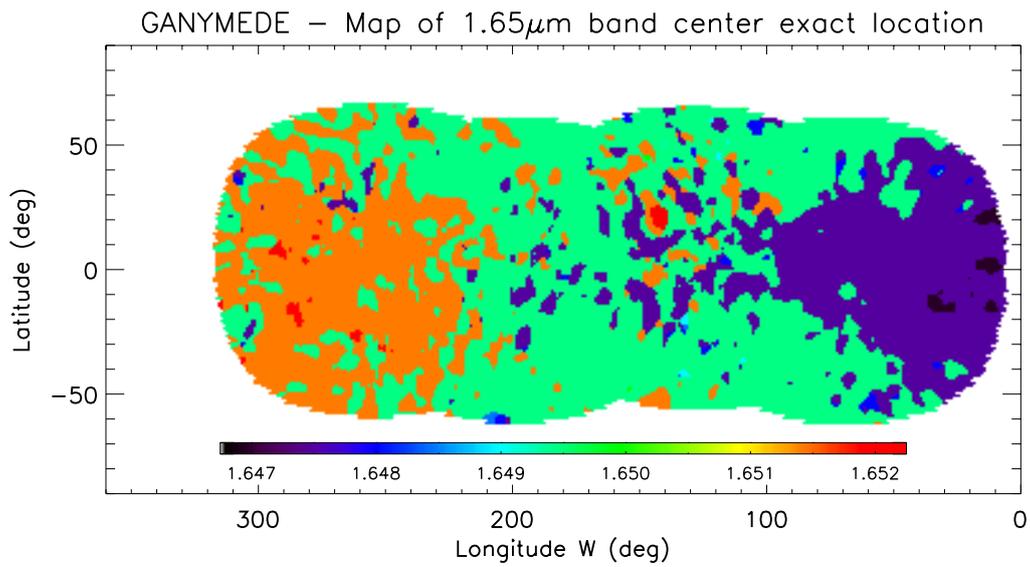

**Figure 7**. Precise location of the 1.65 μm band center over the entire surface of Ganymede. A clear dichotomy, suggesting the influence of the Jovian magnetosphere, is highlighted: most of pixels having a spectrum with a band center ≤ 1.649 μm are on the trailing hemisphere, while most of pixels having a spectrum with a band center ≥ 1.651 μm belong to the leading one. For a majority of pixels, the center of the band is located at 1.6495 μm.

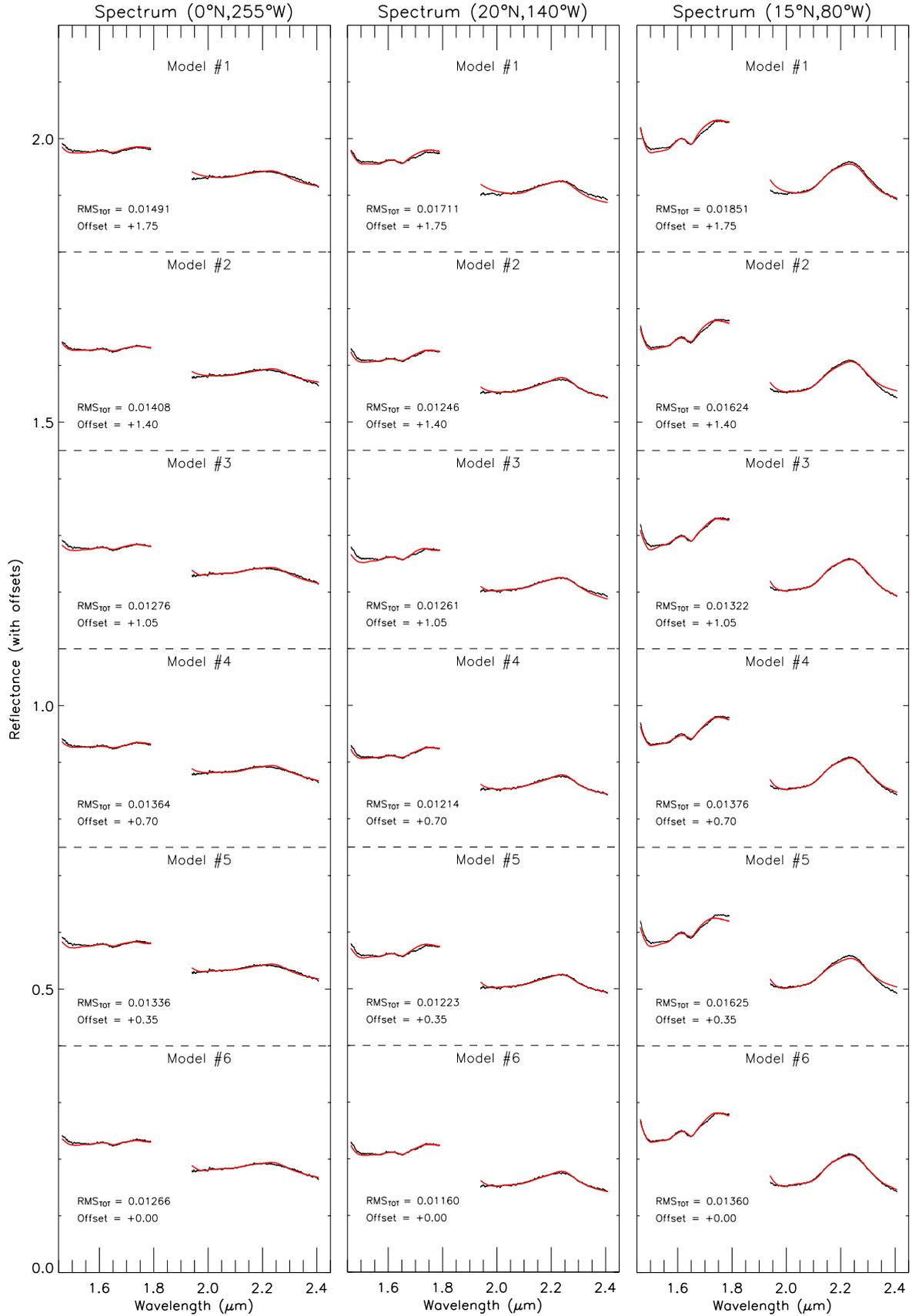

**Figure 8 (1st part)**. Thee measured spectra (black lines) and their modeled spectra (red lines) obtained after performing spectral modeling using the cryogenic spectral library (table 3). Best-fit spectra, i.e. those with the lower $RMS_{TOT}$, are globally obtained for "Model #6". Table 4 summarizes the end-members used and their chemical abundances obtained for each model.

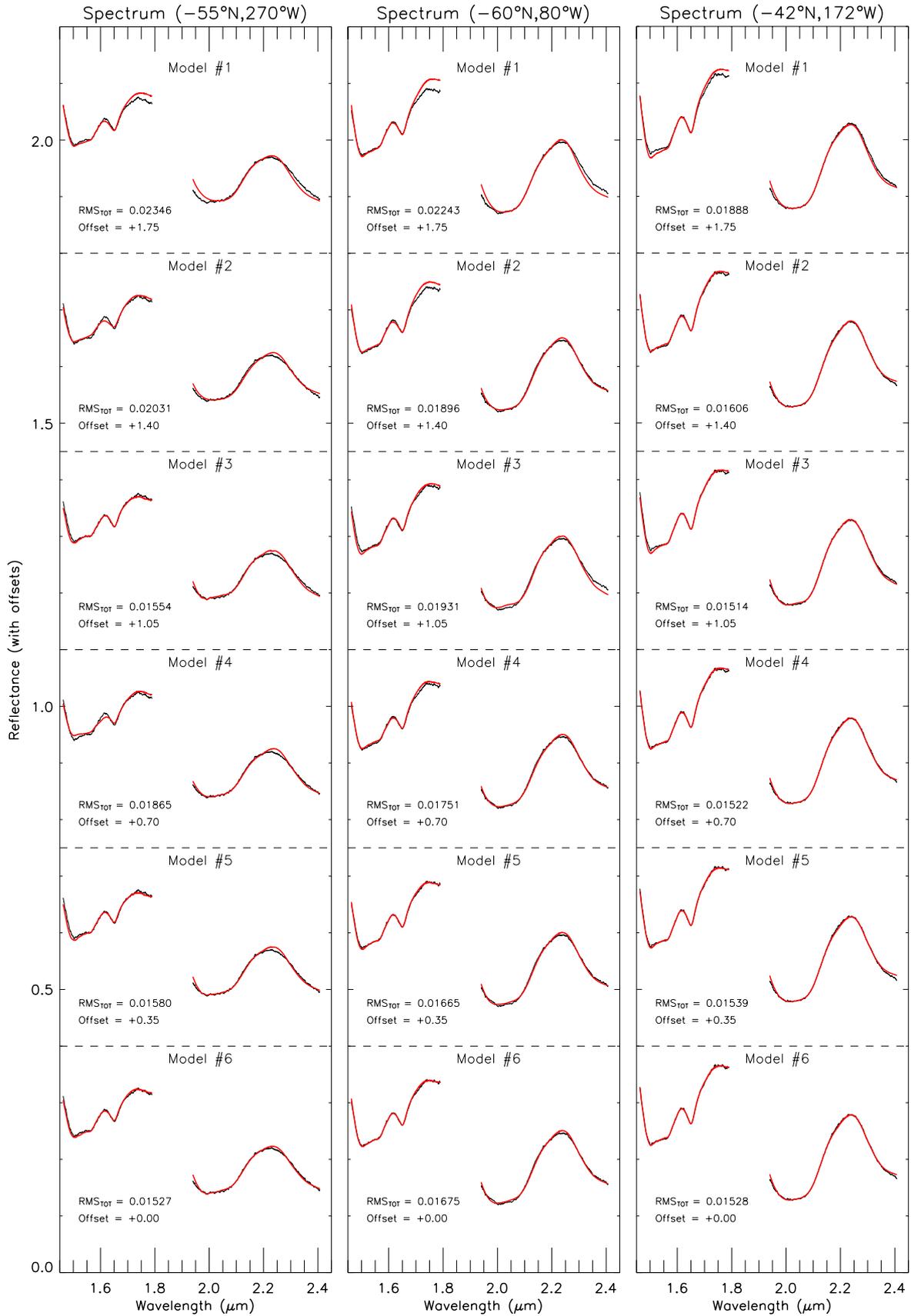

**Figure 8 (2nd part)**. Three others measured spectra (black lines) and their modeled spectra (red lines) obtained after performing spectral modeling using the cryogenic spectral library (table 3). Table 4 provides once again results of the modeling for every model. Finally, with SINFONI's SNR about 1000, error bars on SINFONI's spectra are indistinguishable at the figure's scale.

# Observation Log

| Time (UT) | Target | Exp. (s) | Airmass | Strehl | Obs. long./lat. | S.S.P. long./lat. | $D_{Earth}$ (UA) | $\Phi$ | P.A. |
|---|---|---|---|---|---|---|---|---|---|
| | | | **Observation #1 – 30**[th] **October 2012** | | | | | | |
| 07:05:01 | Ganymede central piece #1 | 10.0 s | 1.449 | 20.3 ± 0.1 | 124.1°W; 3.1°N | 131.1°W; 3.0°N | 4.221 UA | 7.0° | -6.0° |
| 07:06:12 | Ganymede middle-left piece | 10.0 s | 1.449 | 20.7 ± 0.1 | 124.1°W; 3.1°N | 131.1°W; 3.0°N | 4.221 UA | 7.0° | -6.0° |
| 07:07:15 | Ganymede bottom-left piece | 10.0 s | 1.449 | 22.2 ± 0.1 | 124.1°W; 3.1°N | 131.1°W; 3.0°N | 4.221 UA | 7.0° | -6.0° |
| 07:08:24 | Ganymede bottom-middle piece | 10.0 s | 1.450 | 14.7 ± 0.1 | 124.2°W; 3.1°N | 131.2°W; 3.0°N | 4.221 UA | 7.0° | -6.0° |
| 07:09:28 | Ganymede bottom-right piece | 10.0 s | 1.450 | 12.0 ± 0.0 | 124.2°W; 3.1°N | 131.2°W; 3.0°N | 4.221 UA | 7.0° | -6.0° |
| 07:10:37 | Ganymede middle-right piece | 10.0 s | 1.450 | 12.4 ± 0.7 | 124.2°W; 3.1°N | 131.2°W; 3.0°N | 4.221 UA | 7.0° | -6.0° |
| 07:11:40 | Ganymede top-right piece | 10.0 s | 1.451 | 12.7 ± 2.1 | 124.3°W; 3.1°N | 131.3°W; 3.0°N | 4.221 UA | 7.0° | -6.0° |
| 07:12:49 | Ganymede top-middle piece | 10.0 s | 1.451 | 11.3 ± 3.5 | 124.3°W; 3.1°N | 131.3°W; 3.0°N | 4.221 UA | 7.0° | -6.0° |
| 07:13:52 | Ganymede top-left piece | 10.0 s | 1.451 | 10.8 ± 1.2 | 124.4°W; 3.1°N | 131.4°W; 3.0°N | 4.221 UA | 7.0° | -6.0° |
| 07:15:01 | Ganymede central piece #2 | 10.0 s | 1.452 | 09.9 ± 0.0 | 124.4°W; 3.1°N | 131.4°W; 3.0°N | 4.221 UA | 7.0° | -6.0° |
| 07:16:50 | Sky | 10.0 s | 1.453 | / | / | / | / | / | / |
| 07:18:37 | Sky | 10.0 s | 1.456 | / | / | / | / | / | / |
| 07:43:15 | Standard #1 – HD28099 (G2v) | 1.0 s | 1.413 | 20.0 ± 3.0 | / | / | / | / | / |
| 07:53:37 | Standard #2 – HD28099 (G2v) | 1.0 s | 1.440 | 13.8 ± 0.1 | / | / | / | / | / |
| 08:04:37 | Standard #3 – HD25680 (G5v) | 0.83 s | 1.706 | 19.9 ± 1.4 | / | / | / | / | / |
| 08:07:25 | Standard #4 – HD25680 (G5v) | 0.83 s | 1.720 | 22.3 ± 0.1 | / | / | / | / | / |
| | | | **Observation #2 – 23**[rd] **November 2012** | | | | | | |
| 05:59:23 | Standard #1 – HD29714 (G2v) | 4.0 s | 1.510 | 35.5 ± 0.6 | / | / | / | / | / |
| 06:17:25 | Sky | 10.0 s | 1.508 | / | / | / | / | / | / |
| 06:19:00 | Ganymede central piece #1 | 10.0 s | 1.511 | 39.5 ± 0.4 | 253.0°W; 3.1°N | 255.1°W; 2.9°N | 4.077 UA | 2.2° | -7.3° |
| 06:20:17 | Ganymede middle-left piece | 10.0 s | 1.514 | 38.0 ± 0.4 | 253.0°W; 3.1°N | 255.2°W; 2.9°N | 4.077 UA | 2.2° | -7.3° |
| 06:21:42 | Ganymede bottom-left piece | 10.0 s | 1.517 | 37.8 ± 0.5 | 253.1°W; 3.1°N | 255.2°W; 2.9°N | 4.077 UA | 2.2° | -7.3° |
| 06:23:07 | Ganymede bottom-middle piece | 10.0 s | 1.521 | 37.2 ± 0.3 | 253.1°W; 3.1°N | 255.3°W; 2.9°N | 4.077 UA | 2.2° | -7.3° |
| 06:24:23 | Ganymede bottom-right piece | 10.0 s | 1.524 | 38.2 ± 0.3 | 253.1°W; 3.1°N | 255.3°W; 2.9°N | 4.077 UA | 2.2° | -7.3° |
| 06:25:48 | Ganymede middle-right piece | 10.0 s | 1.528 | 36.7 ± 0.8 | 253.2°W; 3.1°N | 255.4°W; 2.9°N | 4.077 UA | 2.2° | -7.3° |
| 06:27:13 | Ganymede top-right piece | 10.0 s | 1.531 | 37.1 ± 0.5 | 253.2°W; 3.1°N | 255.4°W; 2.9°N | 4.077 UA | 2.2° | -7.3° |
| 06:28:31 | Ganymede top-middle piece | 10.0 s | 1.535 | 36.9 ± 0.4 | 253.3°W; 3.1°N | 255.5°W; 2.9°N | 4.077 UA | 2.2° | -7.3° |
| 06:29:56 | Ganymede top-left piece | 10.0 s | 1.539 | 36.8 ± 0.5 | 253.3°W; 3.1°N | 255.5°W; 2.9°N | 4.077 UA | 2.2° | -7.3° |
| 06:31:20 | Ganymede central piece #2 | 10.0 s | 1.543 | 35.1 ± 1.2 | 253.4°W; 3.1°N | 255.6°W; 2.9°N | 4.077 UA | 2.2° | -7.3° |
| 06:33:09 | Sky | 10.0 s | 1.548 | / | / | / | / | / | / |
| 06:35:33 | Sky | 10.0 s | 1.556 | / | / | / | / | / | / |
| 06:43:02 | Standard #2 – HD29150 (G5) | 2.0 s | 1.610 | 33.3 ± 0.2 | / | / | / | / | / |
| | | | **Observation #3 – 17**[th] **February 2015** | | | | | | |
| 04:18:47 | Ganymede central piece #1 | 30.0 s | 1.336 | 24.1 ± 0.4 | 204.2°W; 0.0°N | 202.0°W; 0.1°N | 4.358 UA | 2.2° | +19.5° |
| 04:20:22 | Ganymede middle-left piece | 30.0 s | 1.336 | 23.2 ± 1.1 | 204.3°W; 0.0°N | 202.0°W; 0.1°N | 4.358 UA | 2.2° | +19.5° |
| 04:21:51 | Ganymede bottom-left piece | 30.0 s | 1.337 | 21.1 ± 2.0 | 204.3°W; 0.0°N | 202.1°W; 0.1°N | 4.358 UA | 2.2° | +19.5° |
| 04:23:23 | Ganymede bottom-middle piece | 30.0 s | 1.337 | 20.2 ± 1.4 | 204.4°W; 0.0°N | 202.1°W; 0.1°N | 4.358 UA | 2.2° | +19.5° |
| 04:24:56 | Ganymede bottom-right piece | 30.0 s | 1.338 | 23.4 ± 1.0 | 204.4°W; 0.0°N | 202.2°W; 0.1°N | 4.358 UA | 2.2° | +19.5° |
| 04:26:29 | Ganymede middle-right piece | 30.0 s | 1.338 | 23.2 ± 1.0 | 204.5°W; 0.0°N | 202.3°W; 0.1°N | 4.358 UA | 2.2° | +19.5° |
| 04:27:58 | Ganymede top-right piece | 30.0 s | 1.339 | 21.1 ± 0.3 | 204.5°W; 0.0°N | 202.3°W; 0.1°N | 4.358 UA | 2.2° | +19.5° |
| 04:29:31 | Ganymede top-middle piece | 30.0 s | 1.340 | 21.0 ± 0.1 | 204.6°W; 0.0°N | 202.4°W; 0.1°N | 4.358 UA | 2.2° | +19.5° |
| 04:31:04 | Ganymede top-left piece | 30.0 s | 1.341 | 22.2 ± 0.3 | 204.6°W; 0.0°N | 202.4°W; 0.1°N | 4.358 UA | 2.2° | +19.5° |
| 04:32:35 | Ganymede central piece #2 | 30.0 s | 1.342 | 20.2 ± 2.1 | 204.7°W; 0.0°N | 202.5°W; 0.1°N | 4.358 UA | 2.2° | +19.5° |
| 04:34:13 | Sky | 30.0 s | 1.343 | / | / | / | / | / | / |
| 04:45:07 | Standard #1 – HD111018 (G3v) | 5.0 s | 1.302 | 20.8 ± 0.6 | / | / | / | / | / |
| 04:48:13 | Standard #2 – HD111018 (G3v) | 8.0 s | 1.291 | 18.9 ± 0.1 | / | / | / | / | / |
| | | | **Observation #4 – 08**[th] **March 2015** | | | | | | |
| 01:36:21 | Sky | 30.0 s | 1.425 | / | / | / | / | / | / |
| 01:38:05 | Ganymede central piece #1 | 30.0 s | 1.421 | 24.0 ± 0.9 | 76.6°W; 0.1°N | 70.8°W; 0.0°N | 4.484 UA | 5.8° | +18.9° |
| 01:39:39 | Ganymede middle-left piece | 30.0 s | 1.418 | 20.7 ± 0.1 | 76.6°W; 0.1°N | 70.8°W; 0.0°N | 4.484 UA | 5.8° | +18.9° |
| 01:41:08 | Ganymede bottom-left piece | 30.0 s | 1.415 | 23.2 ± 1.7 | 76.7°W; 0.1°N | 70.9°W; 0.0°N | 4.484 UA | 5.8° | +18.9° |
| 01:42:36 | Ganymede bottom-middle piece | 30.0 s | 1.411 | 26.0 ± 1.6 | 76.7°W; 0.1°N | 70.9°W; 0.0°N | 4.484 UA | 5.8° | +18.9° |
| 01:44:09 | Ganymede bottom-right piece | 30.0 s | 1.408 | 23.5 ± 0.4 | 76.8°W; 0.1°N | 71.0°W; 0.0°N | 4.484 UA | 5.8° | +18.9° |
| 01:45:42 | Ganymede middle-right piece | 30.0 s | 1.405 | 24.7 ± 2.4 | 76.8°W; 0.1°N | 71.1°W; 0.0°N | 4.484 UA | 5.8° | +18.9° |
| 01:47:11 | Ganymede top-right piece | 30.0 s | 1.402 | 22.5 ± 0.1 | 76.9°W; 0.1°N | 71.1°W; 0.0°N | 4.484 UA | 5.8° | +18.9° |
| 01:48:39 | Ganymede top-middle piece | 30.0 s | 1.399 | 18.5 ± 2.0 | 77.0°W; 0.1°N | 71.2°W; 0.0°N | 4.484 UA | 5.8° | +18.9° |
| 01:50:12 | Ganymede top-left piece | 30.0 s | 1.397 | 14.2 ± 0.3 | 77.0°W; 0.1°N | 71.2°W; 0.0°N | 4.484 UA | 5.8° | +18.9° |
| 01:51:45 | Ganymede central piece #2 | 30.0 s | 1.394 | 16.7 ± 0.2 | 77.1°W; 0.1°N | 71.3°W; 0.0°N | 4.484 UA | 5.8° | +18.9° |
| 01:53:23 | Sky | 30.0 s | 1.391 | / | / | / | / | / | / |
| 02:05:46 | Standard #1 – HD103554 (G3v) | 8.0 s | 1.453 | 12.6 ± 0.0 | / | / | / | / | / |
| 09:13:16 | Standard #2 – HD149231 (B3v) | 5.0 s | 1.084 | 22.3 ± 0.7 | / | / | / | / | / |

**Table 1**. Columns from left to right: observation time, observation target, exposure time, airmass, Strehl ratio with standard deviation calculated by the SINFONI RTC, apparent planetographic longitude and latitude of the center of Ganymede as seen by the observer, apparent planetographic longitude and latitude of the sub-solar point ($S.S.P.$) as seen by the observer, distance to Ganymede from Earth ($D_{Earth}$), phase angle ($\Phi$), and North pole position angle ($P.A.$). The last five details are provided by NASA JPL Ephemerides (http://ssd.jpl.nasa.gov/).

**Mosaic reconstruction**

|  | 2012/10/30 | 2012/11/23 | 2015/02/17 | 2015/03/08 |
|---|---|---|---|---|
| **1st piece** | - | - | - | - |
| **2nd piece** | 0.988 | 0.988 | 0.999 | 1.008 |
| **3rd piece** | 0.986 | 0.977 | 0.996 | 0.996 |
| **4th piece** | 0.980 | 0.988 | 0.978 | 0.974 |
| **5th piece** | 0.983 | 0.992 | 0.989 | 0.985 |
| **6th piece** | 1.002 | 1.017 | 1.018 | 0.989 |
| **7th piece** | 1.018 | 1.025 | 1.037 | 1.060 |
| **8th piece** | 1.017 | 1.023 | 1.045 | 1.029 |
| **9th piece** | 1.012 | 1.022 | 1.047 | 1.008 |
| **10th piece** | 1.002 | 1.007 | 1.029 | 0.995 |

**Table 2**. Normalization factors, $f_1$, used for the mosaic reconstruction of every observation. The first piece is, by definition, always the reference used to normalize the nine others. A perfect consistency would consist in $f_1 = 1.000$.

# Cryogenic spectral library

| Chemical Formula | Reference paper | Temperature | Grain size |
| --- | --- | --- | --- |
| $H_2O$ (Crystalline #1) | GHoSST website | 120 K | 10 μm |
| $H_2O$ (Crystalline #2) | GHoSST website | 120 K | 50 μm |
| $H_2O$ (Crystalline #3) | GHoSST website | 120 K | 200 μm |
| $H_2O$ (Crystalline #4) | GHoSST website | 120 K | 1000 μm |
| $H_2O$ (Amorphous #1) | Mastrapa et al. (2008) | 120 K | 10 μm |
| $H_2O$ (Amorphous #2) | Mastrapa et al. (2008) | 120 K | 50 μm |
| $H_2O$ (Amorphous #3) | Mastrapa et al. (2008) | 120 K | 200 μm |
| $H_2O$ (Amorphous #4) | Mastrapa et al. (2008) | 120 K | 1000 μm |
| $H_2SO_4.8(H_2O)$ | Shirley et al. (2010) | 140 K | ~ 50 μm |
| $MgSO_4.6(H_2O)$ | Dalton & Pitman (2012) | 120 K | 100 – 200 μm |
| $MgSO_4.7(H_2O)$ | Dalton & Pitman (2012) | 120 K | 100 – 200 μm |
| $MgSO_4.11(H_2O)$ | Shirley et al. (2010) | 100 K | ~ 50 μm |
| $Na_2SO_4.10(H_2O)$ | Shirley et al. (2010) | 100 K | 100 – 150 μm |
| $Na_2Mg(SO_4)_2.4(H_2O)$ | Dalton & Pitman (2012) | 120 K | 25 – 150 μm |
| $Na_2CO_3.10(H_2O)$ | McCord et al. (2001a) | ~ 100 K | 335 – 500 μm |
| $NaClO_4.2(H_2O)$ | Hanley et al. (2014) | 80 K | / |
| $MgCl_2.2(H_2O)$ | Hanley et al. (2014) | 80 K | / |
| $MgCl_2.4(H_2O)$ | Hanley et al. (2014) | 80 K | / |
| $MgCl_2.6(H_2O)$ | Hanley et al. (2014) | 80 K | / |
| $Mg(ClO_3)_2.6(H_2O)$ | Hanley et al. (2014) | 80 K | / |
| $Mg(ClO_4)_2.6(H_2O)$ | Hanley et al. (2014) | 80 K | / |
| flat dark spectra | / | / | / |

**Table 3**. The cryogenic spectral library publicly available used for the spectral modeling. Temperature and grain size are provided if mentioned in the reference paper. GHoSST website is an online database (https://ghosst.osug.fr) providing crystalline $H_2O$-ice's real and imaginary indices, respectively "*n*" and "*k*", also named optical constants, as a function of wavelength. Mastrapa et al. (2008) also provide these real indices for the amorphous form of $H_2O$-ice. From these indices, reflectance spectra have been obtained for different grain size thanks to the Shkuratov's radiative transfer model (Shkuratov et al. 1999, Poulet et al. 2002).

# Modeling procedure results

| End-members | Pixel (0°N, 255°W) | Pixel (20°N, 140°W) | Pixel (15°N, 80°W) | Pixel (-55°N, 270°W) | Pixel (-60°N, 80°W) | Pixel (-42°N, 172°W) |
|---|---|---|---|---|---|---|
| **Model #1 – Best-fit abundances** | | | | | | |
| **Crystalline H$_2$O-ices** | 10.8 % | 20.9 % | 17.3% | 27.4 % | 73.5 % | 45.8 % |
| **Amorphous H$_2$O-ices** | 6.1 % | 3.7 % | 9.9 % | 18.0 % | 14.0 % | 10.7 % |
| **Darkening agent** | 83.1 % | 75.4 % | 72.8 % | 54.6 % | 12.5 % | 43.5 % |
| **Model #2 – Best-fit abundances** | | | | | | |
| **Crystalline H$_2$O-ices** | 10.8 % | 22.9 % | 14.1% | 17.9 % | 28.8 % | 40.6 % |
| **Amorphous H$_2$O-ices** | 4.0 % | 4.8 % | 13.0 % | 18.4 % | 21.3 % | 11.3 % |
| **Darkening agent** | 75.6 % | 60.2 % | 53.2 % | 40.8 % | 27.8 % | 30.0 % |
| **Sulfuric acid hydrate** | 9.6 % | 12.1 % | 19.7 % | 22.9 % | 22.1 % | 18.1 % |
| **Model #3 – Best-fit abundances** | | | | | | |
| **Crystalline H$_2$O-ices** | 9.8 % | 16.5 % | 16.1% | 29.1 % | 32.1 % | 45.2 % |
| **Amorphous H$_2$O-ices** | 4.3 % | 2.3 % | 4.1 % | 0.4 % | 9.5 % | 3.4 % |
| **Darkening agent** | 78.3 % | 62.3 % | 55.5 % | 37.9 % | 30.1 % | 30.9 % |
| **Sulfates (5)** | 1.0 % | 6.3 % | 20.9 % | 10.9 % | 13.8 % | 10.8 % |
| **Carbonate** | 0.2 % | 8.7 % | 0.0 % | 3.2 % | 1.1 % | 2.1 % |
| **Chlorinated salts (6)** | 6.4 % | 3.9 % | 3.4 % | 18.5 % | 13.4 % | 7.6 % |
| **Model #4 – Best-fit abundances** | | | | | | |
| **Crystalline H$_2$O-ices** | 11.1 % | 23.3 % | 17.9 % | 27.7 % | 31.7 % | 43.0 % |
| **Amorphous H$_2$O-ices** | 3.3 % | 4.5 % | 5.6 % | 7.6 % | 12.3 % | 7.4 % |
| **Darkening agent** | 74.4 % | 56.6 % | 49.4 % | 31.6 % | 25.2 % | 33.4 % |
| **Sulfuric acid hydrate** | 6.2 % | 12.7 % | 11.7 % | 15.2 % | 20.6 % | 11.5 % |
| **Sulfates (5)** | 5.0 % | 2.9 % | 15.4 % | 17.9 % | 10.2 % | 4.7 % |
| **Model #5 – Best-fit abundances** | | | | | | |
| **Crystalline H$_2$O-ices** | 10.7 % | 15.3 % | 12.1% | 28.7 % | 34.1 % | 43.1 % |
| **Amorphous H$_2$O-ices** | 3.0 % | 11.0 % | 14.2 % | 9.9 % | 10.0 % | 11.1 % |
| **Darkening agent** | 76.8 % | 56.8 % | 52.1 % | 39.9 % | 28.8 % | 26.1 % |
| **Sulfuric acid hydrate** | 4.9 % | 11.7 % | 16.8 % | 6.6 % | 18.3 % | 16.6 % |
| **Chlorinated salts (6)** | 4.6 % | 5.2 % | 4.8 % | 14.9 % | 8.8 % | 3.1 % |
| **Model #6 – Best-fit abundances** | | | | | | |
| **Crystalline H$_2$O-ices** | 13.2 % | 23.6 % | 15.2 % | 24.2 % | 33.1 % | 42.0 % |
| **Amorphous H$_2$O-ices** | 1.6 % | 3.3 % | 10.1 % | 14.4 % | 10.5 % | 9.3 % |
| **Darkening agent** | 75.0 % | 56.4 % | 51.7 % | 37.4 % | 27.5 % | 29.7 % |
| **Sulfuric acid hydrate** | 3.6 % | 10.4 % | 9.2 % | 9.5 % | 20.7 % | 15.5 % |
| **Sulfates (2)** | 0.8 % | 3.4 % | 9.5 % | 0.7 % | 0.4 % | 1.7 % |
| *MgSO$_4$.6(H$_2$O)* | *0.5 %* | *2.6 %* | *5.4 %* | *0.7 %* | *0.4 %* | *1.7 %* |
| *Na$_2$Mg(SO$_4$)$_2$.4(H$_2$O)* | *0.3 %* | *0.8 %* | *4.1 %* | *0.0 %* | *0.0 %* | *0.0 %* |
| **Chlorinated salts (2)** | 5.9 % | 2.9 % | 4.3 % | 13.8 % | 7.8 % | 1.8 % |
| *MgCl$_2$.2(H$_2$O)* | *5.2 %* | *1.9 %* | *3.6 %* | *9.6 %* | *4.2 %* | *0.5 %* |
| *Mg(ClO$_3$)$_2$.6(H$_2$O)* | *0.7 %* | *1.0 %* | *0.7 %* | *4.2 %* | *3.6 %* | *0.3 %* |

**Table 4**. Abundances of six best-fit spectra at each step of the modeling procedure. Results highlighted in yellow correspond to the best model (the lowest RMS$_{TOT}$ in figure 8) for a given spectrum.

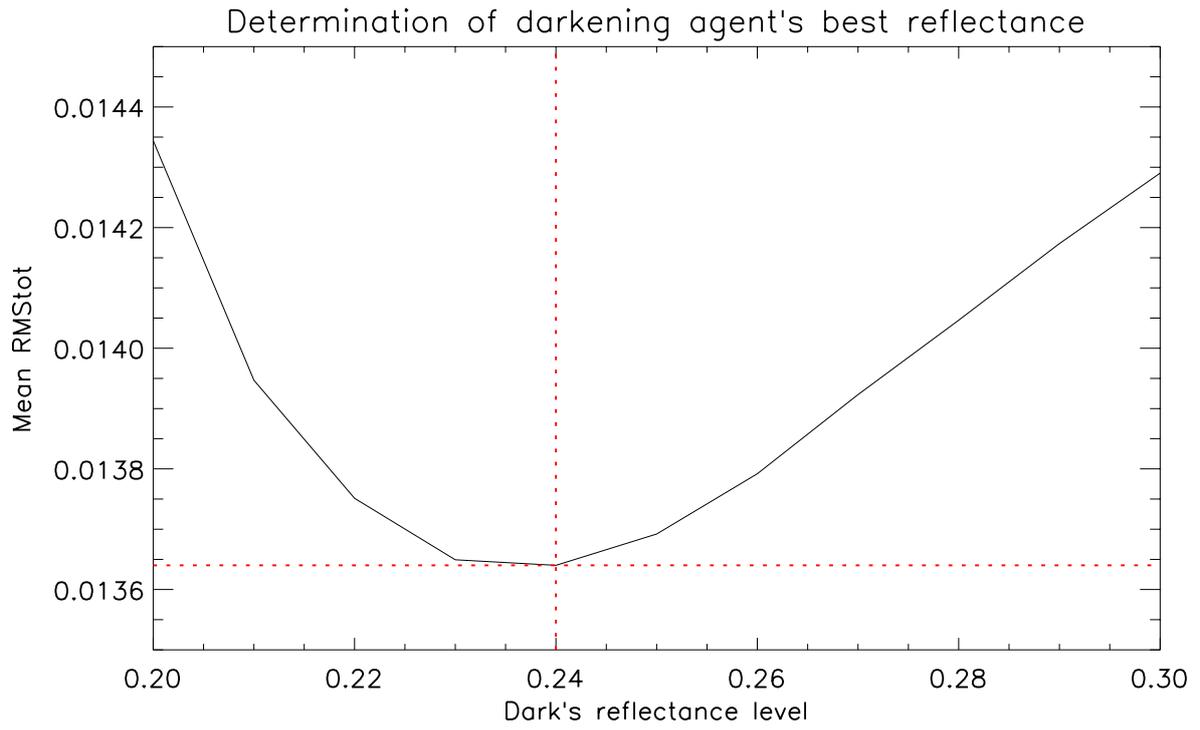

**Figure 9**. Determination of the best reflectance level for the darkening agent to use for the modeling of the whole dataset. The smallest, i.e. best, $RMS_{TOT}$ is obtained when the reflectance level of the dark is equal to 0.24: $RMS_{TOT} = 0.01364$.

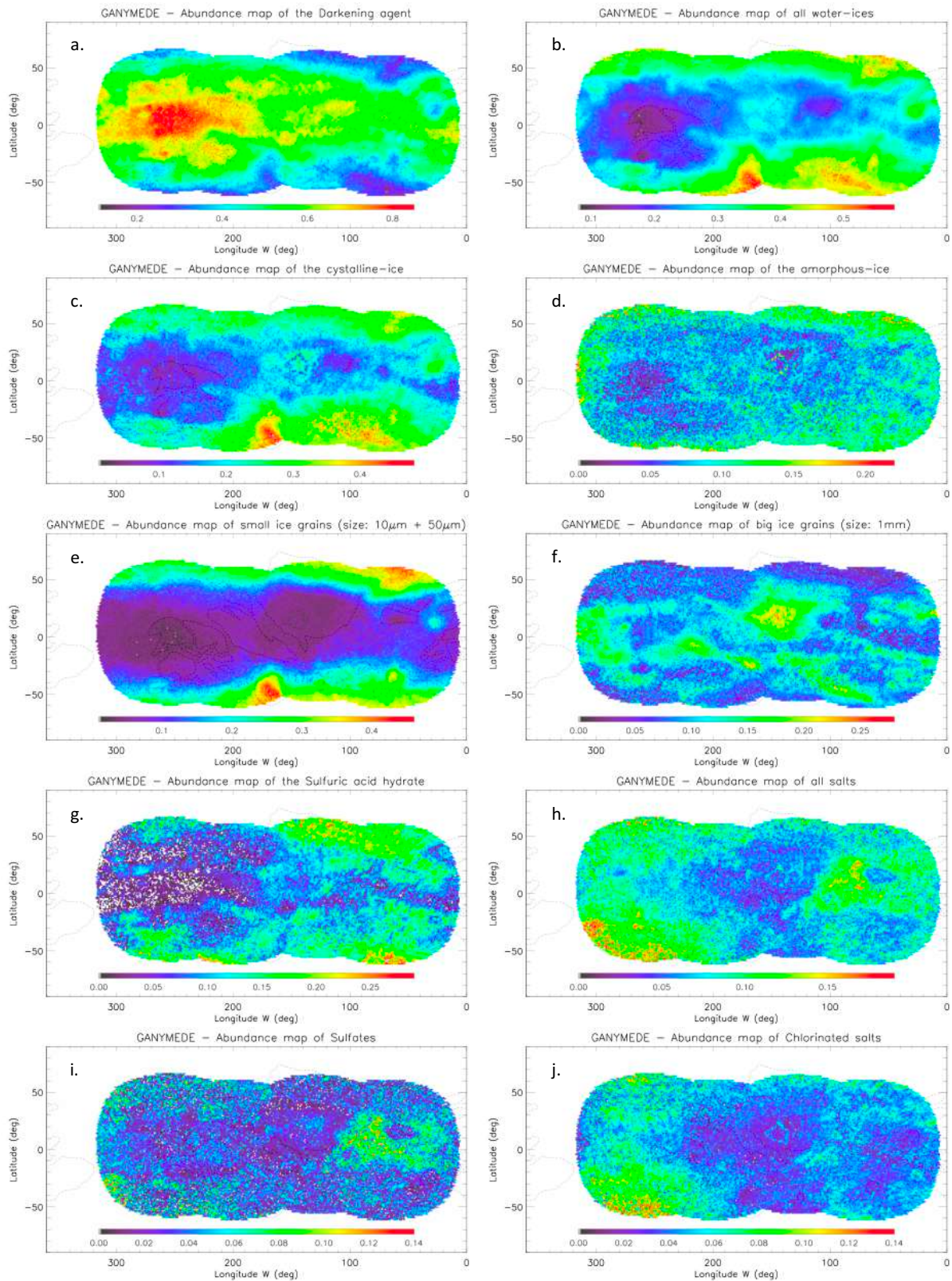

**Figure 10**. Ten different abundance maps of single or multiple end-members used in the final modeling. The darkening agent (a.) is the most abundant end-member of the modeling and is anti-correlated to the whole $H_2O$-ice (b.), but more specifically to its crystalline form (c.) while the amorphous one (d.), less abundant, is distributed quite homogeneously on the surface. The distribution of small grains ($\leq$ 50 μm) of $H_2O$-ice (e.) highlights a strong latitudinal effect, unlike bigger grains (f.). Sulfuric acid hydrate (g.) and salts (h., i., j.) are distributed heterogeneously and could be correlated to endogenous processes. Main geomorphological units delimited thanks to Ganymede USGS map in the visible (figure 4) are superimposed on all the maps.

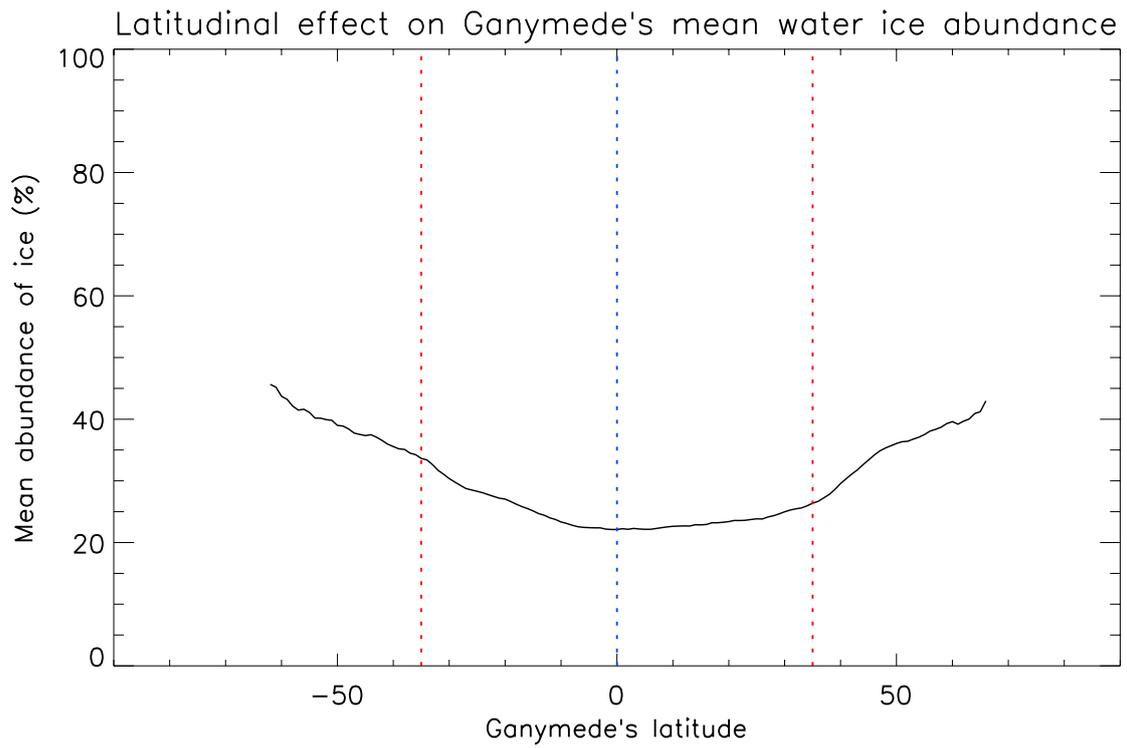

**Figure 11**. Effect of the latitude on the mean abundance of ice on Ganymede's surface. The lowest values are located around the equator (blue dotted lines). A quite important increase is observed around ±35° of latitude (red dotted lines), mostly in the northern hemisphere. This increase coincides with the transition between open and closed field lines of Ganymede's intrinsic magnetic field. This increase in unclear in the southern hemisphere because major bright icy craters (Osiris, Cisti, etc…) are located between −20° and −60° of latitude.

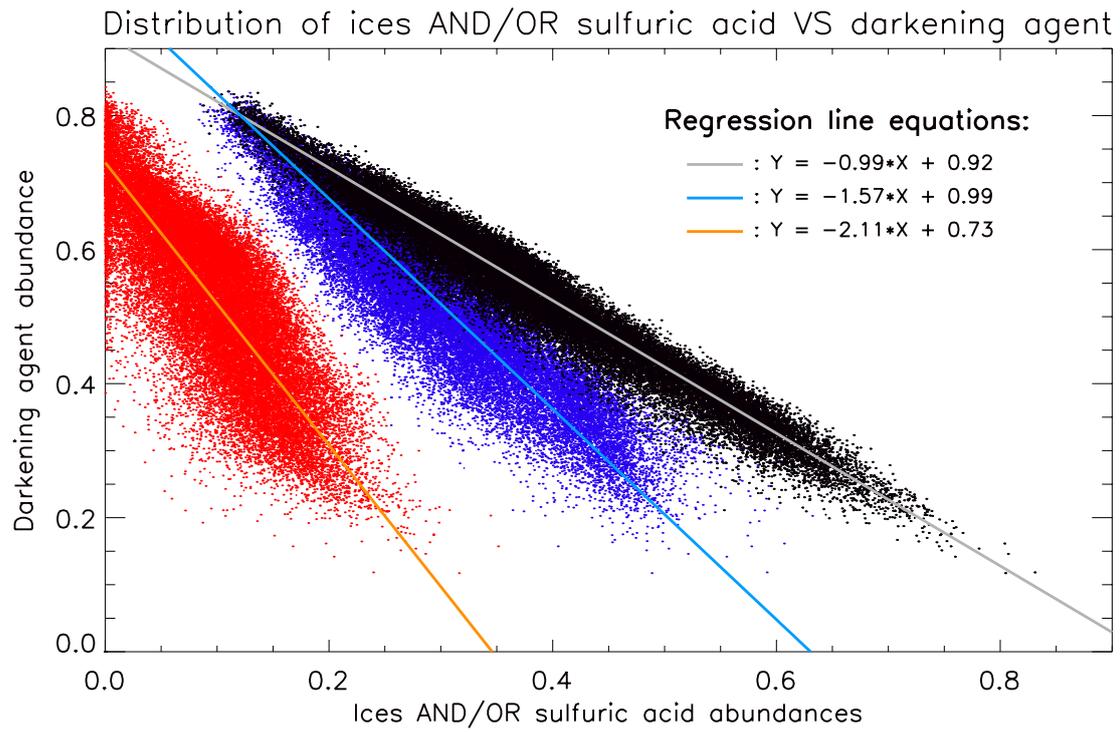

**Figure 12**. Scatter plot showing, for each single pixel, the abundance of the darkening agent as a function of the abundance of the sulfuric acid hydrate (red dots), the total $H_2O$-ice (blue dots) and the sum of the sulfuric acid hydrate and the total $H_2O$-ice (black dots). The clear trends prove the strong spatial anti-correlation between the darkening agent and these two minerals.

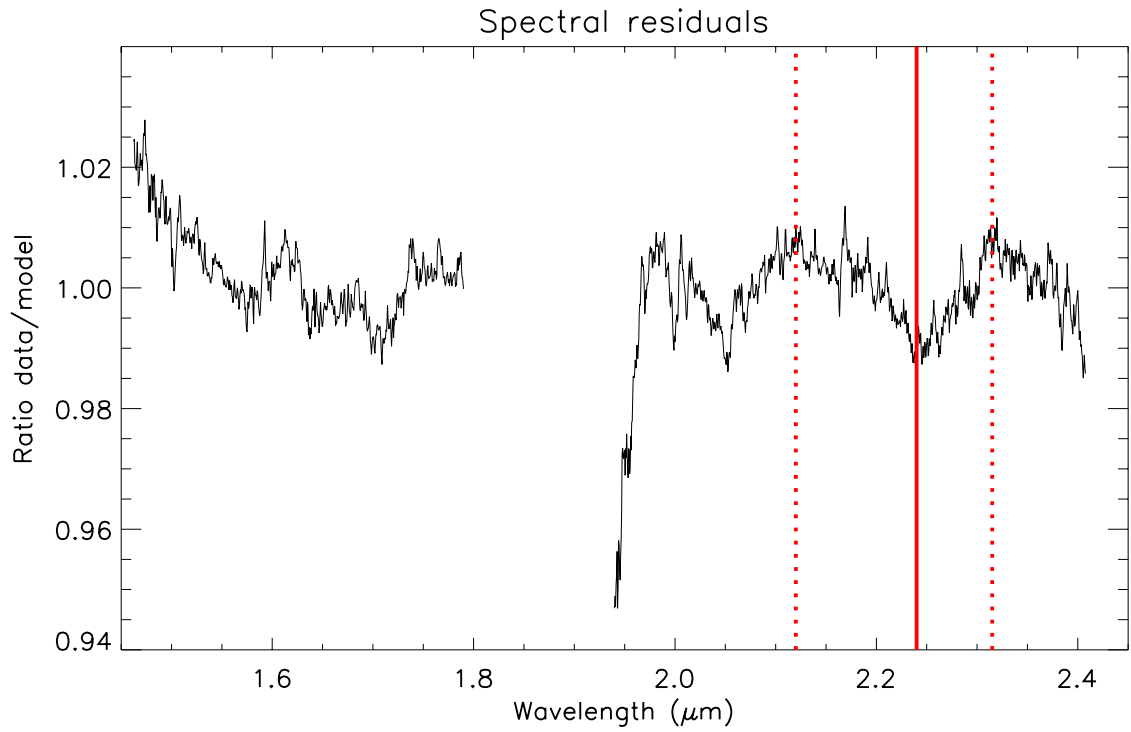

**Figure 13**. Spectral residuals of the modeling. The black line corresponds to the sum of all the spectra divided by the sum of all the modeled spectra. Red dotted vertical lines localize the left and right bounds of a band centered at 2.24 μm (highlighted by the red vertical line) that might be related to an additional absorption feature from a mineral not used during the modeling.

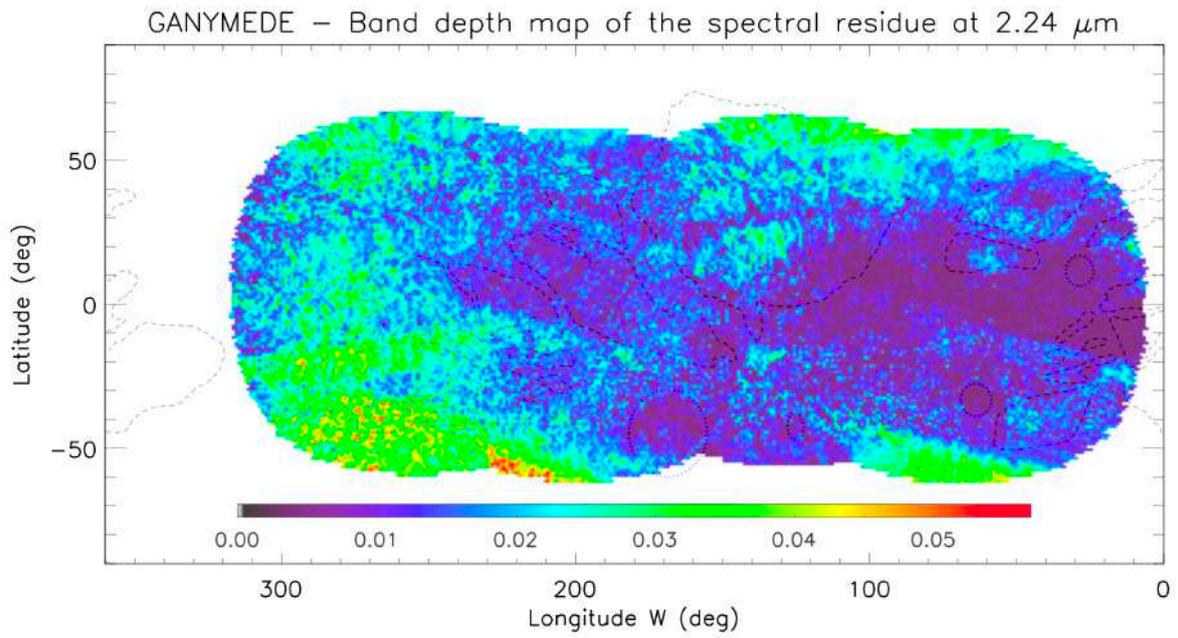

**Figure 14**. Band depth map of the absorption at 2.24 μm observed in figure 12. Overall, the spatial distribution is very similar to the chlorinated salts (figure 10j). Main geomorphological units delimited thanks to Ganymede USGS map in the visible (figure 4) are superimposed on the map.